\documentclass[printer]{aa} 

\usepackage{lineno}
\usepackage{graphicx}
\usepackage{txfonts}
\usepackage{natbib}
\usepackage{color}
\usepackage{hyperref}

\usepackage[normalem]{ulem}

\begin{document}

\title{R(V) parameter of the dust extinction curve: mapping based on hot stars and connections to diffuse interstellar bands}

\author{R. Lallement\inst{1}
\and J.~L. Vergely\inst{2}
\and N.~L.~J. Cox\inst{3}
}

\institute{LIRA, Observatoire de Paris, PSL University, CNRS, 5 Place Jules Janssen, 92190 Meudon, France\\
       \email{rosine.lallement@obspm.fr}
\and 
ACRI-ST, 260 Route du Pin Montard, BP234, 06904 Sophia-Antipolis, France 
\and 
ACRI-ST, Centre d’Etudes et de Recherche de Grasse (CERGA), 10 Av. Nicolas Copernic, 06130 Grasse, France
}

\date{Received ; accepted }
\titlerunning{DIBs and extinction curve}
 
\abstract
{Favored carriers of the Diffuse Interstellar Bands (DIBs) are carbon-based molecules whose constituents have been presumably expelled by dying stars along with those of interstellar dust grains. Some resemblance has been found between sky maps of the $R_\mathrm{V}$ parameter of the extinction curve, governed by grain properties, and those of the 8620  \AA~ (Gaia) DIB abundance relative to extinction density, or DIB strength, but the parallel evolution of DIB carriers and grains and their potential interplay are unknown.}
{Our goal is to compare DIB strengths and $R_\mathrm{V}$ values, either in 3D space using maps or along lines of sight (LOS) toward the hot stars that serve as the sources of individual $R_\mathrm{V}$ measurements and for which DIBs can be easily measured. Independently of the DIBs, these hot stars provide a new partial map of $R_\mathrm{V}$.}
{We assembled $R_\mathrm{V}$ measurements from the literature for the LOS towards 505 hot stars. We computed a new 3D map of extinction density, obtained by inversion of the recently published catalog of extinction data based on Gaia XP spectra and NIR photometry; we used this map to locate the opaque areas responsible for $R_\mathrm{V}$ for each hot star LOS, and we searched for large spatial scale variations of $R_\mathrm{V}$ in this new distribution.  We compared recent 3D maps of local and LOS-integrated $R_\mathrm{V}$ values with the corresponding local or LOS Gaia DIB strengths, calculated from our 3D maps of the DIB and of extinction density.  Using archived spectra, we fitted and measured the 5780, 5797 and 8620  \AA~ DIBs for most of the hot stars with $R_\mathrm{V}$ data, combined them with published data for about 90 stars, then compared the DIB strengths and the $R_\mathrm{V}$ values.}
{$R_\mathrm{V}$s from early-type stars reveal variations on large spatial scales, confirming with entirely independent data the already published 3D pattern. Gaia DIB strength maps show variations at large spatial scale similar to those of $R_\mathrm{V}$ maps, with higher abundance of DIB carriers corresponding to higher $R_\mathrm{V}$s. LOS-integrated Gaia DIB strengths and $R_\mathrm{V}$s based on 3D maps also correlate positively in the $R_\mathrm{V}$ range 2.3-3.4, but there is no connection between the Gaia DIB and $R_\mathrm{V}$ for higher $R_\mathrm{V}$ values. We interpret this absence as due to the disappearance of Gaia DIB carriers in dense clouds containing large grains and with high $R_\mathrm{V}$s, since for LOS crossing such clouds this anti-correlation produces a counteracting effect. Despite strong averaging effects along the LOS to hot stars, DIB and $R_\mathrm{V}$ comparisons reveal significant differences in $R_\mathrm{V}$-DIB relationships among the three DIBs, with an evolution from negative to positive correlation with $R_\mathrm{V}$ from the $\zeta$-type 5797 \AA~, to the $\sigma$-type 5780 \AA~ then 8620 \AA~ DIBs.  Various combinations of DIB strengths or DIB ratios allow to constrain the $R_\mathrm{V}$ range. The relationship between DIB and $R_\mathrm{V}$ seems to depend on how $R_\mathrm{V}$ is defined.}
{3D maps show that variations of environmental conditions on large spatial scale influence both the grains and the 8620 \AA~  DIB. The LOS data show that each DIB has its own specific link between its strength and $R_\mathrm{V}$. The different signs of correlations can be understood in the frame of the recently found bimodal evolution of $R_\mathrm{V}$ with the extinction, with the 8620 (resp. 5797)  \AA~ DIB carrier disappearance in the decreasing (resp. increasing) phase of $R_\mathrm{V}$. The DIB dependence on the $R_\mathrm{V}$ definition suggests that the carrier abundance of each DIB depends on a specific wavelength range of the extinction curve.}


\maketitle

\section{Introduction}\label{intro}

To disentangle the intrinsic spectral properties of astrophysical sources from the effects of interstellar dust, it is necessary to have accurate models of the spectral dependence of these effects, the so-called extinction curve (resp. attenuation) curve for point (resp. extended) sources. The wavelength dependence of the extinction curve (EC) depends on the properties of the dust grains, size distribution and composition \citep{Draine03,Hensley23_astrodust}, and understanding  this spectral dependence and its variability is important to build evolutionary models of the interstellar dust. 
Sophisticated models of the EC have been built. They correspond to families of curves parameterized, among other quantities, by the parameter $R_\mathrm{V}$ equal or close to the inverse of the slope of the extinction curve A($\lambda$)/$A_V$ between the ranges defined by the standard blue and green optical filters, $R_\mathrm{V}$ $\simeq$ $A_V$/E(B-V) \citep{Savage79, Cardelli89, FitzpatrickMassa90, Fitzpatrick99, Gordon23}. Their ranges of application have been extended  from the UV to the optical and MID-infrared, and the descriptions have been improved in several ways to better represent the complexity of the shape and variability of the extinction curve, with the addition of other parameters. The average value for MW objects is $R_V=3.1$ and may reach 5 to 7 for a limited number of stars in or behind very opaque clouds \citep[see, also,][for recent developments]{Xiang17, Fitzpatrick19, MaizApellaniz20, Massa22}. Large values of   $R_\mathrm{V}$ correspond to big grains. Significant progresses have been  recently made on extinction measurements and on the EC. Thanks to the Gaia catalogs of parallaxes, photometric and spectrometric data \citep{Sartoretti18, GaiaDR3_Vallenari, GaiaDR3_Drimmel}, and to photometric and spectroscopic stellar surveys, extinction has been measured for a large and increasing number of stars, and these progresses have been also extended to extinction coefficients \citep[e.g.][]{Danielski18} and to the shape of the EC.  Intermediate-scale and very broad structures (hundreds of  \AA~ wide) in the EC have been uncovered \citep{Massa20, Maiz21, Gordon23, Zhangetal24_LAMOST_CURVE, Gunasekera26}, whose origin is still unknown. For the first time, sky maps of the widely used coefficient $R_\mathrm{V}$ have been obtained \citep{Schlafly17, Zhang_samename23}, and, more recently, a 3D map of $R_\mathrm{V}$ have been computed by \cite{ZhangGreen25} (hereafter ZG25). The 3D map revealed conspicuous variations on large spatial scale (hundred to kilo-parsec size), previously noticed as angular variations in the 2D maps. The most conspicuous trends are the large values of $R_\mathrm{V}$ in the Sagittarius-Carina Arm and the low values in the dense structure call the Split \citep{Lall19}. The authors attributed this strong difference to the existence in Sagittarius-Carina  of a large population of bright early-type stars at the origin of the disappearance of small grains, while the Split has a much smaller number of such stars. Moreover, a remarkable result of ZG25 is the existence of a bi-modal behavior of $R_\mathrm{V}$ with respect to the extinction, namely the existence of a decrease of $R_\mathrm{V}$ with increasing opacity for low to medium extinctions ($A_V$ $\leq$ 2-3 mag), with a minimum of $R_\mathrm{V}$ in translucent-type clouds. This is drastically different from the generally admitted monotonic $R_\mathrm{V}$ increase from low to high density, in response to the increase in size of the dust grains. The authors interpret this as the occurrence of an accretion phase onto the very small grains and a resulting increase of the importance of the newly formed small grains in the total extinction, hence a $R_\mathrm{V}$ decrease. This phase is followed by grain coagulation and strong $R_\mathrm{V}$ increase. Still, physical and chemical processes affecting the dust and at the origin of shape variation and of all features of the extinction wavelength dependence are far from being fully understood and still debated \citep[see, e.g., ][]{ZhangHensley25, Lin25, ZhangGreen25}.

On the other hand, diffuse interstellar bands (DIBs) are the numerous (at least 600)  irregular absorptions detected in the visible and NIR spectra of stars or galaxies \citep[see, e.g. ][for a recent symposium and a review]{IAU14,Cox17}.  Catalogs of DIBs have been produced for the visible domain and near-infrared  \citep[see, e.g.][]{Hobbs09,Fan19,Hamano16}. DIB spectral profiles and wavelength domains point to electronic transitions in large carbonaceous molecules or ultra small carbonaceous grains, and they probably represent a large reservoir of organic matter \citep{Snow2014}.  DIBs are now mapped in the MW \citep{Kos14, Zasowski15, GAIADIB23, Zhao21, Zhao23, Schultheis23_frp, Cox24} and in galaxies at up to 100 Mpc, including LMC, SMC, M31, M33, AM 1353-272, Antennae, NGC6240 outside the Local Group \citep{Heckman2000, Ehrenfreund2002, Cordiner2006, Welty2006, Cordiner11, Monreal15, Monreal18, VanErp25}.  They were also detected in spectra of SNR Ia and quasars \citep[see references in ][]{VanErp25}. These extra-galactic studies show that DIBs are decreasing with the metallicity of the ambient ISM. Except for four bands attributed to the C$_{60}^{+}$ cation \citep{Campbell2016, Cordiner19}, their carriers remain unidentified. The origins of DIB carriers and the processes influencing their evolution and abundance are still a matter of debate. Do they all (or some of them) originate from the ejecta of AGB stars and supernova remnants, transformed in the interstellar medium and selected based on their resistance to UV radiation and shocks? Do they all (or some of them) originate from the intense chemical reactions occurring in opaque clouds, and are they then re-injected at the periphery of these clouds into star-forming regions? Is their disappearance—well observed in dense molecular cores—due to their accretion onto dust grains—in which case the grains and DIB carriers evolve in parallel—or is it due to a change in their charge state?

Links  between DIB strengths and the shape of the EC have been searched for, mostly dedicated to the UV range \citep{Witt83,Benvenuti89,Krelowski92, Desert95,Xiang11,Xiang17,Lall25}, with a recent extension to the visible \citep{Lall24}. Reasons for existing links with the UV EC are of various types. PAHs are among favored potential carriers of DIBs and also believed to be sources of the well-known, conspicuous $\simeq$ 2175  \AA~ absorption excess (the so-called UV bump). Observed correlations between the wavelength shifts of the UV bump center and those of some IR emission bands attributed to PAHs reinforce a connection between DIB carriers,  PAHs and UV bump \citep{Blasberger17}. For this reason, DIB carrier identifications would help understanding the origin of the bump.  Another reason for such searches, as mentioned above,  is a better knowledge of the parallel evolution of dust particle size distributions from very diffuse gas to opaque molecular clouds on the one hand, and the disappearance of DIB carriers of various families on the other hand. As a matter of fact, for all DIBs,  EWs cease to follow the amount of extinction where lines of sight are crossing dense areas, and, instead, are strongly decreasing in UV-shielded molecular clouds, the so-called skin effect. While it has independently been well established that DIB strengths depend on the amount of UV radiation \citep{Vos11}, and, as a result, varying ionization degrees of the carriers could be the unique source of the skin effect, it seems that the situation is more complex. The relationships between DIB EWs and columns of HI and H$_{2}$ are markedly different from one DIB to the other \citep{Vos11, Lan2015, Fan17, Ensor17}. Indeed, \cite{Ensor17} has shown that a factor other than the UV field significantly influences the intensity of DIBs. There is also some evidence that two DIBs sharing about the same sensitivity to the radiation field decrease differently in cloud cores \citep{Elyajouri2017b}.  \cite{BerLall17} suggested that one of the sources of the DIB carrier disappearance could be the accretion of DIB carriers onto grains in transition regions between diffuse and dense clouds. If this is the case, for at least some of the DIBs, their strengths should have some connection to the grain size distribution, hence to the parameters of the EC.

The results of past searches for a DIB-EC connection were sometimes positive, sometimes negative, often unclear \citep[see][for recent extensive analyses]{Xiang11, Xiang17}, and no clear consensus has been reached. One difficulty is that tight correlations between LOS measurements of DIB EWs and other quantities can be excluded, due to the integration along the path to the target and the frequent crossing of several clouds which may be of different nature. Even in the mono-cloud LOS case, because clouds are inhomogeneous, the way the LOS crosses the outskirts or the central part does affect partly the averaged properties. Secondly, because DIB equivalent widths (EWs) depend at first order on the integrated amount of interstellar matter, EWs of a unique DIB can not be used as diagnostics of the physico-chemical properties of the absorbing medium, instead EWs normalized to the amount of absorbing matter (i.e., the extinction or the reddening) are the relevant quantities, which makes necessary simultaneous measurements of DIBs and extinction. An exception is the use of DIB ratios, requiring measurements of at least two different DIBs. Those two difficulties are now on the verge of being overcome.  Massive amounts of individual extinction data are now available and can be used to construct three-dimensional (3D) maps of the dust extinction and it is possible to use those maps to estimate the integrated extinction along any line of sight and compute the extinction- (resp. reddening-) normalized  DIB, or DIB strength,  i.e. EW/$A_V$ (resp. EW/E(B-V)). Moreover, the new availability of 3D maps of dust extinction allows to distinguish mono-cloud type lines of sight from those characterized by  multiple clouds  distributed over distance. Based on such a selection, it has been made possible to show that some strong optical DIBs are connected to the height, the width, and the position of the 2175 \AA~ UV bump of the extinction curve \citep{Lall24, Lall25}. Recently, \cite{Saydjari26} analyzed Gaia BP/RP spectra in conjunction with DIBs in high-resolution SDSS/APOGEE spectra and in the Gaia radial Velocity Spectrometer (RVS) spectra, and showed that inter-correlations between EC-features of all large- and medium-wavelength scales also include the DIBs.

\begin{figure*}
\includegraphics[width=0.8\textwidth]{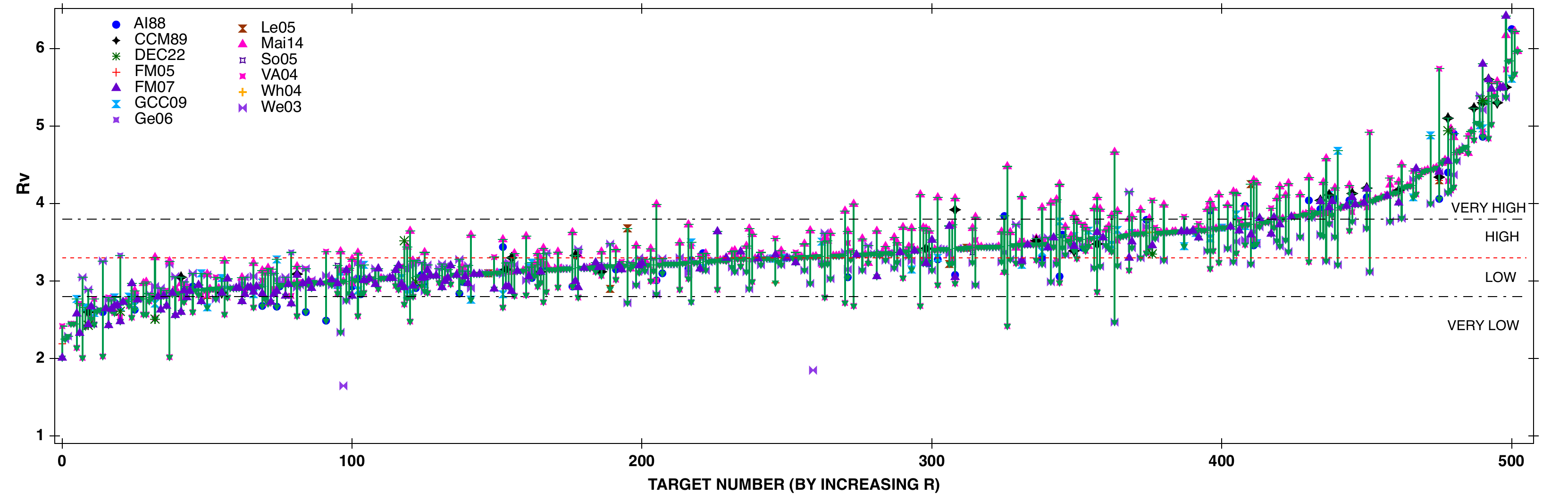}
\caption{$R_\mathrm{V}$ of target stars with fitted DIBs: $R_\mathrm{V}$ values from different sources and the same star are superimposed. The vertical green segments join the minimum and maximum values in the case of more than one estimate. Targets are ordered by increasing average Rv. Ai88: \cite{Aiello88}; CCM89: \cite{Cardelli89}; FM05: \cite{Fitzpatrick05}; FM07: \cite{Fitzpatrick2007}; GCC09: \cite{Gordon09}; Ge06: \cite{Geminale06}; Le05: \cite{Lewis05}; Mai14: \cite{Maiz14}.; So05: \cite{Sofia2005}; Va04: \cite{Valencic04};   We03: \cite{Wegner03}; Wh04: \cite{Whittet04} ; }
\label{graph_allrvs}
\end{figure*}

\begin{figure}
\includegraphics[width=0.45\textwidth]{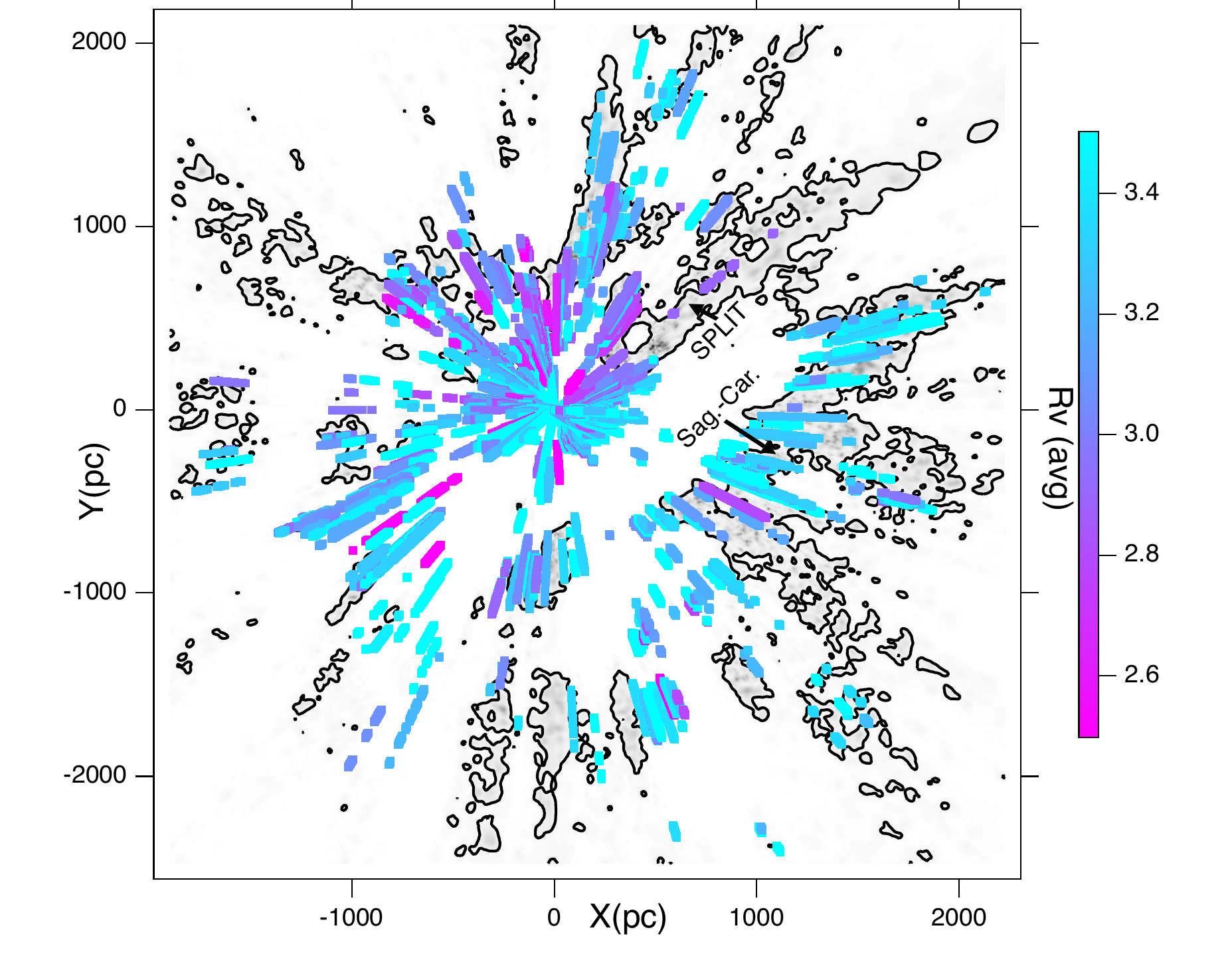}
\caption{Large scale variations of Rv:  $R_\mathrm{V}$ value (color scale) for individual lines of sight to hot stars distributed in areas of high extinction (see text), superimposed on iso-extinction density contours for a value of 0.002 mag pc$^{-1}$ in the 3D map described in section 3.2. $R_V$ values below 2.5 (resp. above 3.5) are given the colors of 2.5 (resp. 3.5). Targets located in the Split are only a few (see text).}
\label{Large_scale_rv}
\end{figure}

\begin{figure*}
\includegraphics[width=0.8\textwidth]{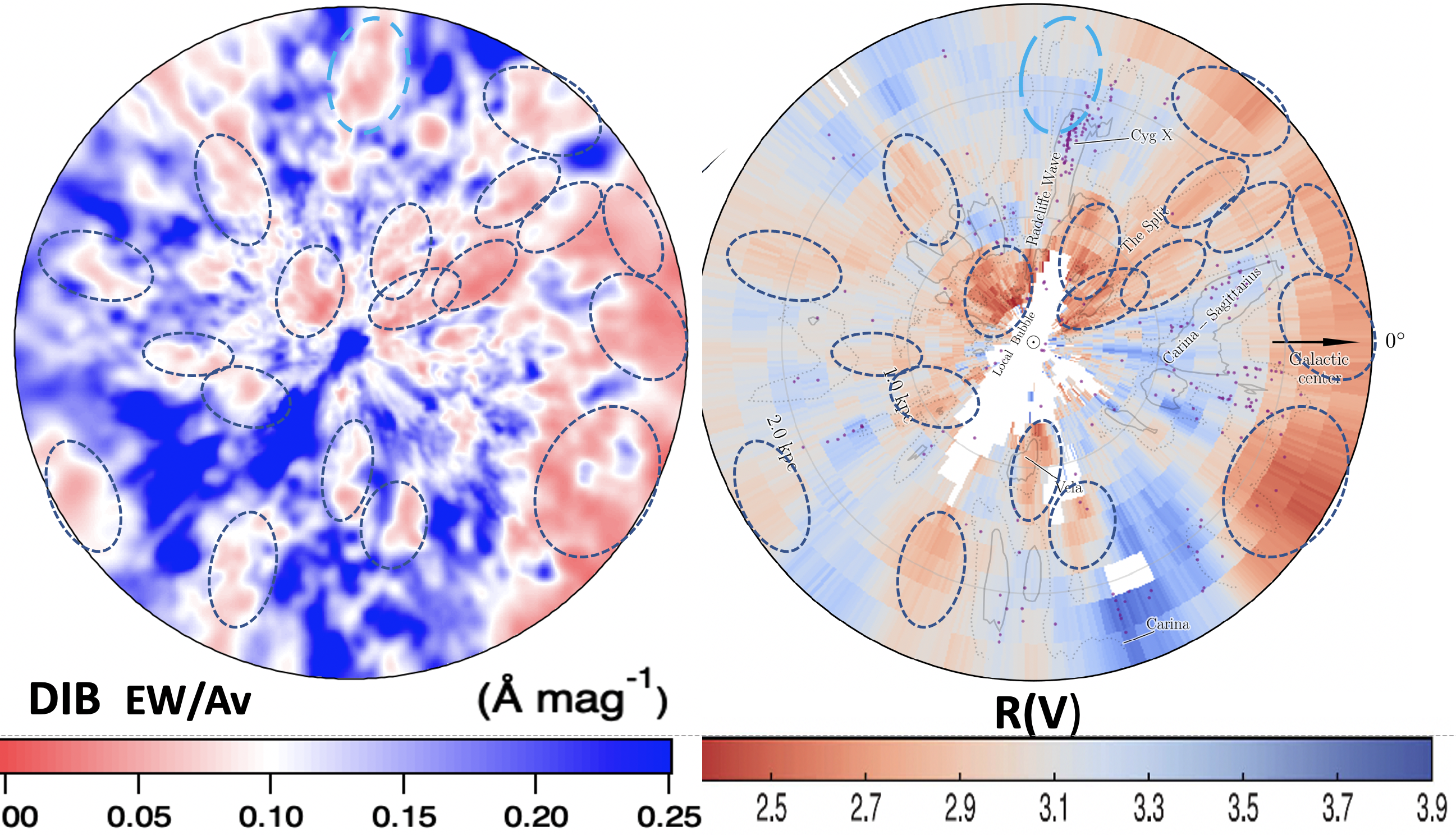}
\caption{Large scale variations of local DIB strength and local $R_\mathrm{V}$ and : Left: 8620 \AA~ DIB strength averaged from -200 to +200 pc along the z direction and projected onto the Galactic Plane. Superimposed ellipses illustrate regions of low DIB strength (red color). Right: reproduction of Fig. 1 from ZG25 showing $R_\mathrm{V}$, with identical superimposed ellipses. Low DIB strength correspond to low Rv. A counterexample  can be seen in the first quadrant (green ellipse).}
\label{Large_scale_rv_dibtoext}
\end{figure*}

\begin{figure}
\includegraphics[width=0.45\textwidth,height=5cm]{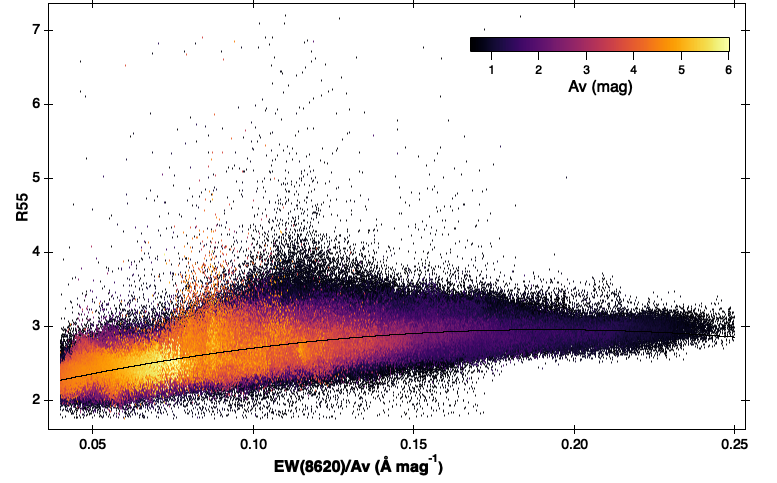}
\includegraphics[width=0.45\textwidth,height=7cm]{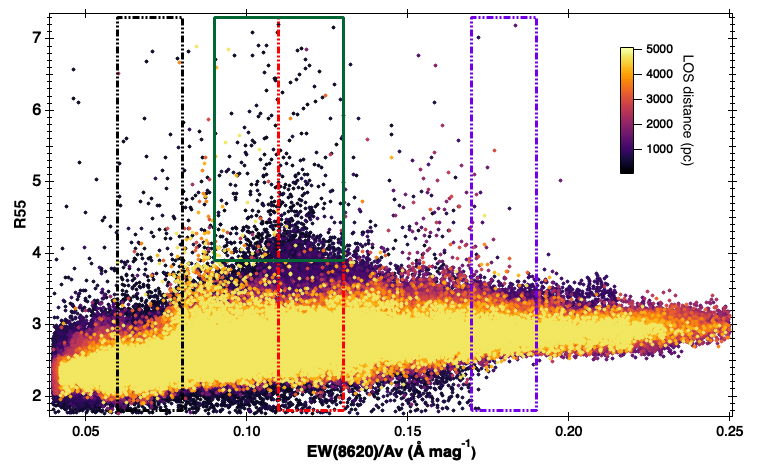}
\caption{Top: ZG25 LOS R55 values and Gaia DIB strength from the \cite{Cox24} DIB carrier and extinction density maps. The figure is restricted to $A_V$$\geq$0.6 mag to avoid large uncertainties on the DIB to extinction. There is a positive correlation between R55 and the DIB and a flattening at large DIB strength. The color code, with high extinctions at the foreground, reveals that decreasing R55's correspond in average to an increase in extinction, a trend discovered by ZG25 for several nearby clouds and found here for a majority of LOS probing the diffuse to translucent medium. A series of discrepant very high R55 points is visible and a wide bump of higher than average R55 data points is also discernible under those outliers (see text). The black line is a polynomial fit of order 5 to the whole set of data points. Bottom: same as top except for the color code, here showing the LOS distance, with high distances at the foreground. The four rectangles delimitate the regions analyzed in Fig. \ref{histos}.}
\label{R55GAIAMAPS}
\end{figure}

\begin{figure}
\includegraphics[width=0.42\textwidth]{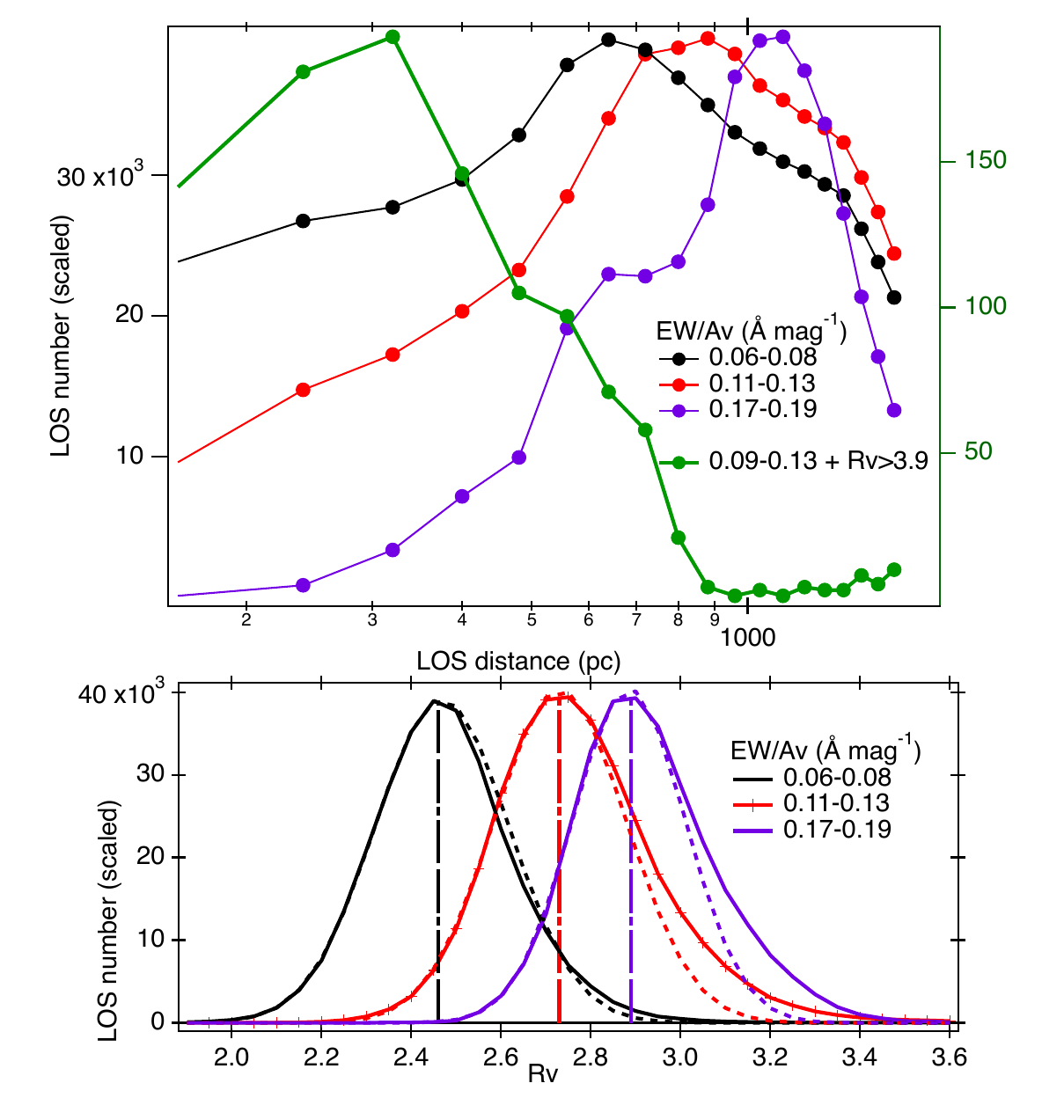}
\caption{Top: Histograms of LOS distances for the four regions in Fig. \ref{R55GAIAMAPS} scaled to reach the top of figure.  Left scale: low, medium, high DIB strength (black, red, violet lines respectively). Right scale: the very high R55 region (green curve). Bottom: Histograms of R55 for the low, medium, high DIB strength regions (black, red, violet lines respectively). The histogram for the very high R55 region is not shown. Gaussian fits to the pre-maximum parts of the distributions are shown as dashed lines, to illustrate the appearance of extended wings of large values (see text).}
\label{histos}   
\end{figure}

This work is extending the previous work presented in \cite{Lall24, Lall25} and is focusing on the links between DIBs and the $R_\mathrm{V}$ parameter. Section 2 is devoted to the description of the sources of individual $R_\mathrm{V}$ measurements for early-type stars, and to the comparison between the large scale pattern of $R_\mathrm{V}$'s revealed by the 3D map of ZG25 and the one of these individual data. It describes the building of a new 3D extinction map based on the same extinction catalog as the one used for the 3D map of $R_\mathrm{V}$ of ZG25 and the use of this new map to locate the source regions of the individual Rv's. Section 3 describes comparisons between $R_\mathrm{V}$ and the Gaia DIB based on their 3D maps. Section 4 describes the sources of DIB data for the stars possessing measured $R_\mathrm{V}$s and the building of the new catalog of 5780, 5797, and 8620 \AA~ DIB measurements based on archive spectra or published catalogs . Section 5 describes relationships between $R_\mathrm{V}$s and DIB strengths for these hot star data. The results are summarized and discussed in section 6.

\section{Rv parameter from individual catalogs}

\subsection{Sources of Rv}


For our compilation, we started with the work of \cite{Xiang17} (hereafter X17). In this study devoted to the link between DIBs and the UV part of the extinction curve, the authors extracted $R_\mathrm{V}$ values from 10 different sources. We kept their 95 early-type target stars and their $R_\mathrm{V}$ values listed in X17, then extended our search to targets from the sources of $R_\mathrm{V}$ quoted in X17, but absent from their compilation, and to targets from other $R_\mathrm{V}$ catalogs. We retained targets with archive spectra suitable for DIB measurements. For a large number of targets there are several determinations of $R_\mathrm{V}$ and we kept all of them.  Table \ref{tabstarsshort}  contains the various estimates of $R_\mathrm{V}$ and the corresponding sources. Some sources correspond to extensive catalogs while others contain one or a few targets. Figure \ref{graph_allrvs}  illustrates the similarity, and also the differences among the various $R_\mathrm{V}$ determinations. It also shows that $R_\mathrm{V}$'s below $\simeq$2.8 (hereafter designed as very low) and above $\simeq$3.8 (hereafter designed as very high) are much less numerous than medium values between 2.8 and 3.8. This is partly due to the large distance to the O and B target stars used to study the EC and averaging effects. For the medium values, we will call low (resp. high) the first and second halves of the series, i.e. $2.8\leq R_\mathrm{V}\leq 3.3$ (resp. $3.3\leq R_\mathrm{V}\leq 3.8$. However, note that the R55 parameter from ZG25 used in section \ref{3D_LOS} is not exactly $R_\mathrm{V}$. We have excluded of the following analyses the two outliers with $R_\mathrm{V}$ lower than 2, clearly visible in the figure. Table  \ref{tabstarsshort}  contains the determinations of $R_\mathrm{V}$ and of the extinction $A_\mathrm{V}$ and the sources. The O, B stars chosen for EC study are for a major part significantly extinguished. 

\subsection{Building a new 3D extinction map}

In order to check whether the individual LOS $R_\mathrm{V}$ data based on precise spectroscopic data confirm the large scale variability detected by ZG25, we need to locate in space the obscuring dust clouds at the origin of the extinction and mostly responsible for the LOS $R_\mathrm{V}$ value. To be as much as possible consistent as possible with the ZG25 $R_\mathrm{V}$ map, we took advantage of our developed technique of hierarchical  inversion \citep[][for a description]{Lallement22, Vergely22} and applied it to the \cite{Zhang23} catalog of extinctions. \cite{Zhang23} combined Gaia BP and RP spectra with a large set of photometric data, and devised a machine learning technique trained on LAMOST high resolution spectra to produce a catalog of stellar and extinction parameters for 220 millions of stars. The same dataset was further used by ZG25 to build 3D maps of the $R_\mathrm{V}$ parameter extending up to 5 kpc from the Sun. 
For our newly computed extinction map, the spatial resolution in the inner regions close to the Sun is reaching 5 pc in most areas, while the resolution in the external regions is degraded down to 50 pc depending on the target spatial density. We show in the Appendix A1 a comparison between the inner part of the map along the Galactic plane and the corresponding map at very high spatial resolution  obtained by \cite{Edenhofer24} based on the same data, and limited to 1.2 kpc. We can see from the figure that in this distance range, all structures are similar to those present in the high resolution map of \cite{Edenhofer24}, although slightly more extended due to our lower resolution.  Iso-contours of the dense structures extending up to 2.6 kpc from the Sun can also be seen in Figure \ref{Large_scale_rv}. Our full map, extending to 10 kpc, and associated tools are available at the ESA Datalabs platform \footnote{\url{https://datalabs.esa.int/}}.

\begin{table*}
\caption{Extinction properties. Sources are (1): \cite{Aiello88}; (2): \cite{Decleir22}; (3): \cite{Fitzpatrick2007}; (4): \cite{Gordon09}; (5): \cite{Maiz14} ;  (6): \cite{Geminale06}; (7): \cite{Cardelli89}; (8): \cite{Valencic04}; (9): \cite{Wegner03};  (10): \cite{Whittet04} ;  (11): \cite{Lewis05}; (12): \cite{Sofia2005};  (13):   \cite{Fitzpatrick05};  (X) \cite{Xiang17}. Averaged values of the extinction Av and of Rv in case of multiple sources are reported in the DIB table. For several targets from (3), an imposed value R=3.1 has been used and is not taken into account in the average. The full table is available at the CDS.}
\label{tabstarsshort}
\begin{tabular}{lllll}
\hline
star&lII,bII&d(pc)&$AV$ sources&R sources\\
\hline
HD161061&0.53,0.58&1729&3.01(9)&3.04(9)\\
HD 162978 &4.54,0.3&964&1.2(8)1.06(9)1.29(5)1.27X&3.46(1)3.54(8)3.64(9)4.3(5)\\
HD 164536&5.96,-0.91&1526&0.93(3)1.09(8)0.76(9)1.03(5)&3.73(3)3.45(9)4.22(5)\\
Herschel36&5.97,-1.17&1290&4.62(5)&5.57(5)5.3(7)\\
HD 315033 &6,-1.34&1153&1.56(3)1.44(8)&4.45(3)4.12(8)\\
HD 164794&6.01,-1.2&1250&1.18(9)1.32(5)1.22X&4.06(1)3.57(9)4.2(5)\\
HD 315032 &6.02,-1.29&1379&1.01(3)1.1(8)&3.61(3)3.94(8)\\
HD 315035&6.05,-1.47&1225&1.75(3)&3.97(3)\\
HD164906&6.05,-1.33&1344&2.17(4)1.72(9)&5.01(4)5.39(9)\\
HD 164816 &6.06,-1.2&1291&0.95(3)0.99(4)1.09(8)0.85(9)1.05(5)&3.26(3)3.21(4)3.52(8)3.26(9)4.09(5)\\
\hline
\end{tabular}
\end{table*}

\subsection{Large spatial scale variations from individual Rv data and comparison with the Zhang and Green 3D map}

For each LOS for which at least one $R_\mathrm{V}$ is available, we used our new 3D extinction density map described above to compute the contribution to the target star extinction of all segments of the LOS. In practice, we decomposed each LOS in 1 pc long segments and computed the extinction A(r) of each segment, as well as its ratio  f(r)=A(r)/A(Rstar) to the cumulated extinction at the target star distance r=Rstar. We then selected all parts of the LOS where f(r)>10$^{-3}$ and plotted all the corresponding LOS points, which has the effect of enhancing those regions that contribute in a major way to the extinction. We colored these fractions of the LOS, i.e. regions where $R_\mathrm{V}$ is generated, based on the value of the measured $R_\mathrm{V}$. Figure \ref{Large_scale_rv} shows these series of LOS fractions, superimposed on an iso-extinction density contour for a value of 0.002 mag pc$^{-1}$ in the 3D extinction density map. As expected, colored fractions of the LOS coincide with the most opaque areas. We can see that, despite some small scale variability, and the fact that we apply a unique $R_\mathrm{V}$ to each LOS, there are visible large scale trends. E.g., similarly to the Fig. 1 of ZG25 reproduced in our Fig. \ref{Large_scale_rv_dibtoext}, the Sagittarius-Carina dense areas have a higher $R_\mathrm{V}$ compared to other regions, while the Split has low $R_\mathrm{V}$ values. The Split trend is less extensively mapped than in the ZG25 figure, owing to the scarcity of bright O, B stars in the Split, a scarcity invoked by ZG25 as the origin of small grain survival and lower Rv. This similarity reinforces the surprising large scale pattern of $R_\mathrm{V}$ discovered by ZG25, here seen based on a O,B star $R_\mathrm{V}$ catalog.

\section{3D maps of Rv and of the Gaia DIB strength}\label{3D}


In this section we use the 3D maps and associated catalogs of ZG25 and compare with the information contained in the 3D map of the Gaia DIB. ZG25 estimated the {\it spectroscopic} parameter R55 defined as A(550nm)/(A(440nm)-A(550nm), tightly related to Rv. The average relationship quoted by the authors is $R_\mathrm{V}$ = 1.1 R55 + 0.07. ZG25 made available two types of 3D distributions, cumulative (LOS integrated) and local (differential) values of R55. In what follows, if we use data from the catalogs, we use the R55 values, however in the discussion and comparisons with DIBs and other sources of $R_\mathrm{V}$ we generally use the term Rv. 

The 3D map of local Gaia DIB strength, which measures how the DIB carrier abundance compares with the dust grain opacity in a specific region, is derived from both DIB and extinction 3D distributions presented in \cite{Cox24}, by simply dividing the former by the latter. The DIB carrier 3D distribution has been obtained by inversion of the catalog of DIB LOS EW measurements of \cite{GAIADIB23}. The 8620 \AA~ DIB is fortunately present in the wavelength range of the RVS on board Gaia. The extinction map used to normalize the DIB carrier density to the opacity (grain) density has been obtained by inversion of extinction values for the same catalog of stars used for the DIB map.  As explained by \cite{Cox24}, the choice of a 3D extinction density map computed in the same way and based on the same targets than those used to produce the DIB map ensures the same spatial resolution and the same limitations for the two maps and a meaningful ratio of the two quantities.

\subsection{Local values of Rv and Gaia DIB strength: large spatial scale variations}

Figure \ref{Large_scale_rv_dibtoext} allows a visual comparison between the $R_\mathrm{V}$ distribution found by ZG25 and the Gaia DIB strength distribution from \cite{Cox24}. It is associating side by side a colored map of the local Gaia DIB strength within 2.4 kpc from the Sun and a reproduction of the Figure 1 from ZG25 showing  $R_\mathrm{V}$  in the same area.  Both images are not exactly Galactic plane values. The DIB strength map was obtained by averaging the DIB to extinction ratio along the z axis from -200 to +200 pc. The ZG25 figure is obtained by averaging A(440nm) and A(550nm) along the vertical z-axis between -400 and +400 pc and restricted to high quality measurements and targets with E(B-V) above 0.1 mag.  In order to facilitate the comparison and guide the eye, we have superimposed ellipses on the low DIB areas and reported the same ellipses on the $R_\mathrm{V}$ map. Despite the completely independent data sets and the different ways to obtain a Galactic plane projection, it is clear that there are important similarities between the $R_\mathrm{V}$ and DIB strength patterns. In the majority of cases, regions of low DIB strength correspond to low Rv. This strongly suggests that the processes regulating the DIB carrier abundance have a connection with the grain size distribution characteristics. It is important to note that in both maps the quantities are averaged over relatively large volumes (in both cases with volumes increasing with distance to the Sun), and for this reason the contribution of very high density cloud cores is small. For the $R_\mathrm{V}$ map, this explains why high values of $R_\mathrm{V}$ do not appear. For the DIB map, this implies a dominant role of the diffuse ISM. As we will discuss in next sections, this dominant role is also increased by the large abundance of Gaia DIB carriers relative to grains in low density ISM and the disappearance of the carriers in the molecular phase.

\subsection{Line-of-sight Rv and Gaia DIB equivalent width normalized to the extinction}\label{3D_LOS}


For what follows,  we made use of the R55 values from the ZG25 catalog appropriate for the entire LOS (from the Sun to the cell center) instead of the local (differential) value discussed in the previous section. We used this dataset and our 3D maps of the Gaia DIB and of extinction to investigate further the relationships between LOS $R_\mathrm{V}$ values and LOS DIB strengths in the low to moderate extinction regime. In \cite{Lall24} we already showed that the Gaia DIB strength is increasing with $R_\mathrm{V}$ in the range $A_V$$\leq$2-3 mag. To do so we had used the $R_\mathrm{V}$ catalog of \cite{Schlafly17} and the Gaia DIB strength map. Here,  we extracted from the ZG25 catalog all defined values of LOS-integrated R55 and the associated characteristics of the LOS, i.e. the distance and the coordinates of the LOS extremity. We then used them to integrate from the Sun to the LOS extremity within both the Gaia DIB 3D map of \cite{Cox24}  and within the corresponding 3D extinction map mentioned above based on the same targets  than those used for the DIB map. The ratio of the two integrals is the normalized DIB (or DIB strength) for each LOS. We caution that due to the low resolution of these maps, extinction may be diluted in larger volumes than actual extensions of the clouds and extinction values may be underestimated. Because low values of the DIB EW and/or of the extinction result in uncertain DIB strength, we restricted our study to LOS integrated extinction above 0.6 mag, which corresponded to DIB EWs above $\simeq$0.02 \AA~ and DIB strengths above $\simeq$0.03 \AA~ per mag. The resulting dataset is made of $\simeq$ 2.5 million LOS with an average length of 1.9 kpc.

 In figure \ref{R55GAIAMAPS} we show how R55 varies as a function of the LOS-averaged DIB strength. A  positive correlation is found for the large majority of the LOS, and corresponds to R55 variations between 2.2 and 3.3. There is a flattening of the R55-DIB dependence for large values of the DIB strength. The color scale in the top graph shows that the low values of R55  correspond to high values of the LOS extinction, a trend similar to the one found by ZG25 for the four nearby clouds. A conspicuous group of data points at very high R55 is visible for DIB strengths below $\simeq$ 0.2  \AA~ mag$^{-1}$.  An examination of the LOS distances for these points (bottom graph of figure \ref{R55GAIAMAPS}) shows that they all correspond to nearby clouds. A high value of the LOS-averaged $R_\mathrm{V}$ is indeed expected here only for low distance cells. As a matter of fact, for extended LOS, the cloud multiplicity is erasing the individual properties of the clouds and the diffuse phase which covers large fractions of the LOS has a strong influence.

Fig. \ref{R55GAIAMAPS} shows a kind of wide bump for the medium values of the DIB strength (say, between $\simeq$ 0.07 and  $\simeq$  0.18  \AA~ mag$^{-1}$). The curvature of the high order polynomial fit to the data is reflecting this pattern. In order to investigate further the characteristics of the bump we selected data points in four regions delimited by rectangles in Fig. \ref{R55GAIAMAPS}. For each region we show in Figure \ref{histos} the histograms of LOS distances and those of  R55. The very high R55 area (green rectangle) corresponds, as said above, to nearby clouds, and LOS crossing the dense parts of these clouds. The region with rather high DIB strength (violet rectangle) corresponds to very distant LOS (histogram maximum for 1.2 kpc LOS)  and the quasi-constant average R55 value around 3. The region with the lower DIB strength (black rectangle) corresponds to shorter distances (maximum around 600 pc). The central region (red rectangle) corresponds to intermediate distances (maximun (asymetric) at 1 kpc).  We fitted the R55 histograms to a Gaussian, restricting the fit to the left part of the histogram, i.e. values lower than the value at the top of the histogram (Fig. \ref{histos} bottom). There are increasing discrepancies between this Gaussian curve and the actual histogram for increasing R55, with ubiquitous increasing wings of high values. The existence of these skewed distributions of R55 for a given DIB strength range shows that other processes than those responsible for the global increase are at play. Our interpretation of the whole pattern is that LOS of increasing length have an increased probability to cross dense clouds filled with large grains, which has the effect of increasing the LOS-averaged R55 and produce the skewing of the R55 distribution, because large grains are producing a flatter wavelength extinction curve and larger R55. The presence of those clouds (and the resulting increase of R55) has no effect on the LOS-averaged DIB strength, because DIB carriers are disappearing in opaque regions.  As a result, and somewhat unfortunately, it is not possible to use LOS-integrated values of this DIB to detect high R clouds. As we will see in the next section, other DIBs may be more useful.

\section{DIB measurements in spectra of early-type stars with extinction curve information} \label{dib_extraction}

We have assembled 5780, 5797 and 8620  \AA~ DIB data for the target stars of Table \ref{tabstarsshort}, i.e. targets for which we could find at least one published estimate of the extinction curve parameter Rv. The choice of the 5780 and 5797 \AA DIBs is motivated by their strengths, hence maximum data availability. The 8620  \AA~ DIB is exceptional due to its central wavelength comprised in a spectral range of prime interest in terms of stellar atmospheric parameter constraints. \cite{Munari08} were the first to make extensive measurements of this DIB based on the RAVE survey. The wavelength interval of the RVS spectrometer on board Gaia was also chosen for this reason and is at the origin of the largest number of target stars with 8620  \AA~ DIB measurements. The association of the 5780 and 5797  \AA~ DIBs is also especially interesting since they are known to represent different classes of DIBs with respect to the evolution of the DIB to extinction ratio from diffuse to dense media. More specifically, the decrease of the DIB 5780 for lines of sight crossing increasingly opaque regions occurs at a smaller opacity compared to one of the DIB 5797, and the differences between such strength variations are believed to be due to their different behaviors with respect  to the UV radiation field \citep[see, e.g.][]{Vos11}. For this reason, DIB 5780 and DIB 5797 have been classified as respectively members of the $\sigma$ and $\zeta$ families of DIBs \citep{Krelowski87}. On the other hand, based on SDSS spectra, \cite{Lan2015} found that their strengths vary in a similar way in low and very low opacity lines of sight.  The carriers of DIB 8620, on the other hand, have been found to be among the most abundant with respect to dust in low density parts of clouds \citep{Lall24}, a behavior confirmed by the large scale height of the carriers above the Galactic plane, as shown by \cite{Zhao23}. 

The 5780 and 5797 DIB data have the following sources: for 91 stars we extracted EWs directly from the extensive compilation of \cite{Xiang17} (hereafter X17), devoted to the comparison between DIBs and the extinction curve in the UV. For targets without values from this catalog,  we have searched for appropriate spectra in the ESO Portal archive and in the Polarbase online facility. From ESO we downloaded spectra recorded with the UVES (166 stars), FEROS (37), ESPRESSO (4), XSHOOTER (18), HARPS (2)  spectrographs. From Polarbase we downloaded spectra recorded with the CFHT ESPAdOnS (101 stars) and Pic du Midi NARVAL (30 stars) spectro-polarimeters. In this latter case, when the polarimetric mode was the only one available, we use the total intensity V.  For a few cases we used spectra in one of the Stokes components. However, DIBs being unpolarized, as shown by \cite{Cox2011b}, this does not alter the absorption. For 5 stars we used measurements available from the \cite{Fan19} catalog. In this case, and for the 5780 \AA~ DIB we summed the EW's of the two adjacent absorptions separated by 0.95 \AA~ and distinguished by \cite{Fan19}.  The following sources were used for the 8620 \AA~DIB:  UVES (123), ESPAdOnS (115), NARVAL (13), XSHOOTER (27), FEROS (18), \cite{Fan19} catalog (5) , GIRAFFE (2). For one star the EW is taken from the Gaia archive data presented in \cite{GAIADIB23}. The more limited number of data for the Gaia DIB is mainly due to the wavelength limitation of several spectrographs in the NIR. All EW's are listed in Table \ref{tabdibsshort} along with the average values of Rv and Av.

We fitted individually the three DIBs in the classical way: -1) definition of the continuum through the selection of wavelength intervals around the DIB free of detectable lines and of artefact due to Cosmic Ray impact on the detector or imperfect transitions between consecutive orders in the reduction process; -2) fitting of the continuum to a polynomial of order 2 to 4 depending on the shape of the continuum; -3) division of the spectrum by the fitted continuum; -4) computation of the EW between two limits visually chosen, ensuring that the wavelength interval is larger than the known interval for the mono-cloud DIBs and smaller than this interval augmented by the average kinematical broadening in the nearby ISM. Examples of typical fits and the wavelength intervals used for continuum and DIB are shown in Figure \ref{dib-fits} from the Appendix. 
For almost all stars, the quasi-totality of the uncertainty on the EW is linked to the definition of the continuum, and not to the signal-to-noise (S/N), because the targets are bright early-type stars and the S/N is relatively high, often above 100. Estimating the error due to the choice of the continuum is very difficult. For the 5780 \AA~ DIB, there are only very weak stellar lines and the relative uncertainty is less than 10\%. In the case of the 5797 \AA~ DIB, the proximity with a broad stellar line red ward of the DIB makes the continuum fitting sometimes more uncertain depending on the velocity difference between the ISM and the target star and the stellar type (see Fig. \ref{dib-fits}). The relative error $\delta$(EW)=$\sigma$(EW)/EW may reach 30\%. In the case of the weak and shallow Gaia DIB, the main difficulty comes from its location in the far red and the frequent existence at those wavelengths of fringes or irregularities inherent to the spectrographs. As a result, there are strong differences from one instrument to the other. Many of the FEROS measurements have a large uncertainty above 40\% and have been rejected. UVES measurements are all useful and characterized by $\delta$(EW) between 5 and 30\%. ESPAdOnS spectra at those wavelengths are quite regular and $\delta$(EW) is often smaller than 15\%. Our goal being to search for global trends, and because multi-cloud aspects are averaging and erasing the trends, we have estimated uncertainties based on the used spectrograph and the DIB for DIBs at 5797 and 5780 \AA~ . For the Gaia DIB all errors were estimated visually. We kept all measurements for which the relative error is less than 40\%.

\begin{table*}
\caption{DIB measurements (mA, milliAngstr\"om). INS1 refers to measurements of the 5780 and 5797Å DIBs. INS2 refers to measurements of the Gaia
DIB. E: ESPADONS; EO: ESPRESSO; F: FEROS; Fn: Fan et al. (2019) catalog; Ga: Gaia; H: HARPS; N: NARVAL; R: GIRAFFE; U: UVES;
X: Xiang et al. (2017) catalog; XS: XSHOOTER. Av and Rv are the average value of the various determinations for the same star. The full table is available at the CDS.}
\label{tabdibsshort}
\begin{tabular}{lllllllllll}
\hline
star&EW5780&sig&EW5797&sig&EW8620&sig&INS1&INS2&Av&Rv\\
\hline
&mA&mA&mA&mA&mA&mA&&&mag&\\
\hline
HD161061&497&23&120&11&350&18&XS&XS&3.01&3.04\\
HD 162978 &210&19&59&9&103&23&X&U&1.2&3.73\\
HD 164536&136&16&30&8&97&21&U&U&0.95&3.8\\
Herschel36&440&21&107&11&125&12&U&XS&4.62&5.44\\
HD 315033 &129&17&56&10&67&15&U&U&1.5&4.29\\
HD 164794&154&8&39&2&108&27&X&U+F&1.24&3.94\\
HD 315032 &149&16&53&10&100&22&U&U&1.06&3.78\\
HD 315035&172&16&63&10& & &U&&1.75&3.97\\
HD164906&153&6&65&5& & &EO&&1.95&5.2\\
HD 164816 &156&9&42&5&55&13&E&E&0.99&3.47\\
\hline
\end{tabular}
\end{table*}

\section{Rv and DIB relationships based on early-type stars}

\begin{figure*}
\includegraphics[width=0.90\textwidth,height=7cm]{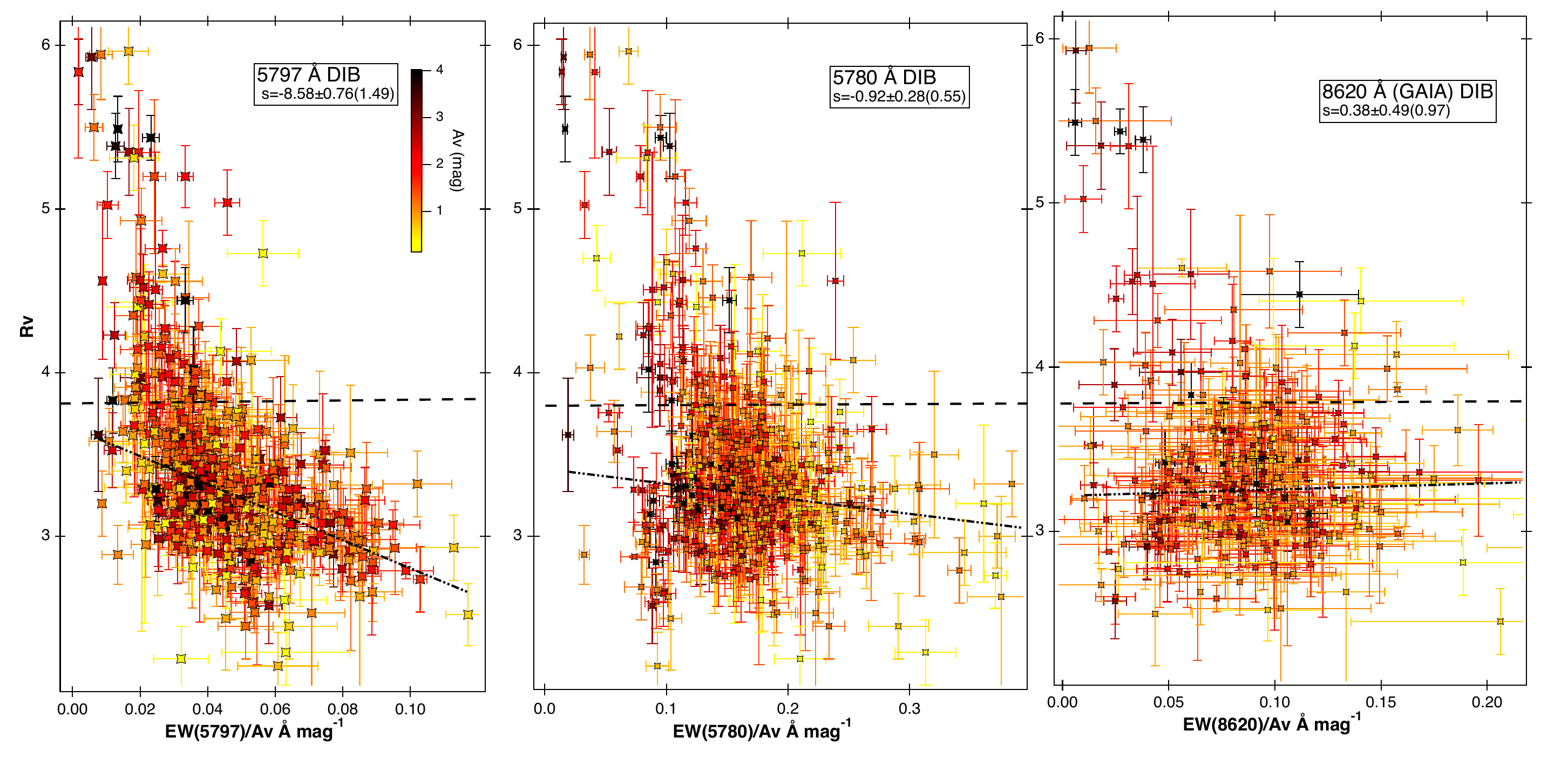}
\caption{Average $R_\mathrm{V}$ from individual LOS and DIB EW's normalized to the extinction $A_V$. Indicated slopes correspond to linear fits with data errors derived from the dispersion around the fit (dot-dashed lines), for $R_\mathrm{V}\leq3.8$ (below the black dashed line). The uncertainty on the slope and the associated value in parentheses correspond to 65 (resp. 95) \% confidence levels.} 
\label{NORMAV_3DIBS_ALL}  
\end{figure*}

Figure \ref{NORMAV_3DIBS_ALL} shows $R_\mathrm{V}$ averaged over the sources as a function of the DIB EWs normalized to the extinction. Error bars on $R_\mathrm{V}$ are half of the maximum difference between two measurements for the same star, or set to 0.2 in case of a unique determination. Error bars on the normalized DIB are deduced from the uncertainties on the EWs defined in the previous section, and assume that the error on the extinction, or on the reddening, is negligible in comparison with the one on the DIB EW. There is a strong dispersion for all three DIBs. We know that this is expected, even in the presence of a strong link acting locally,  due to the use of LOS-integrated measurements and of mostly distant early-type stars  (see table \ref{tabstarsshort}). Such LOS often cross several groups of clouds, and averaging effects are present. For all three DIBs, very high $R_\mathrm{V}$'s (say, above 5) correspond to very low DIB strengths. For lower $R_\mathrm{V}$'s, the results are very surprisingly different among the DIBs. Linear fits were performed for data points with $R_\mathrm{V}\leq3.8$. For those fits errors bars on the fit coefficients were derived by iteration from the dispersion of data points around the linear fit. One can see that there is an anti-correlation between $R_\mathrm{V}$ and the 5797 \AA~ DIB, but that the anti-correlation disappears for the 5780 \AA~ DIB and is replaced by a hint of a positive correlation for the Gaia DIB. The surprising marked difference between the three DIBs shows that the abundance of their carriers is linked to different types of grain distributions, affecting differently the EC in the optical domain and Rv. In the case of the GAIA DIB, we unexpectedly found that the relationship we derived from the 3D maps, i.e., an increase of  $R_\mathrm{V}$ for increasing DIB strength, is apparently absent in the case of these individual data. One of the reasons is certainly the multi-cloud aspect of most LOS and the strong averaging effects. Another reason will be discussed below.

\begin{figure*}
\includegraphics[width=0.90\textwidth,height=7cm]{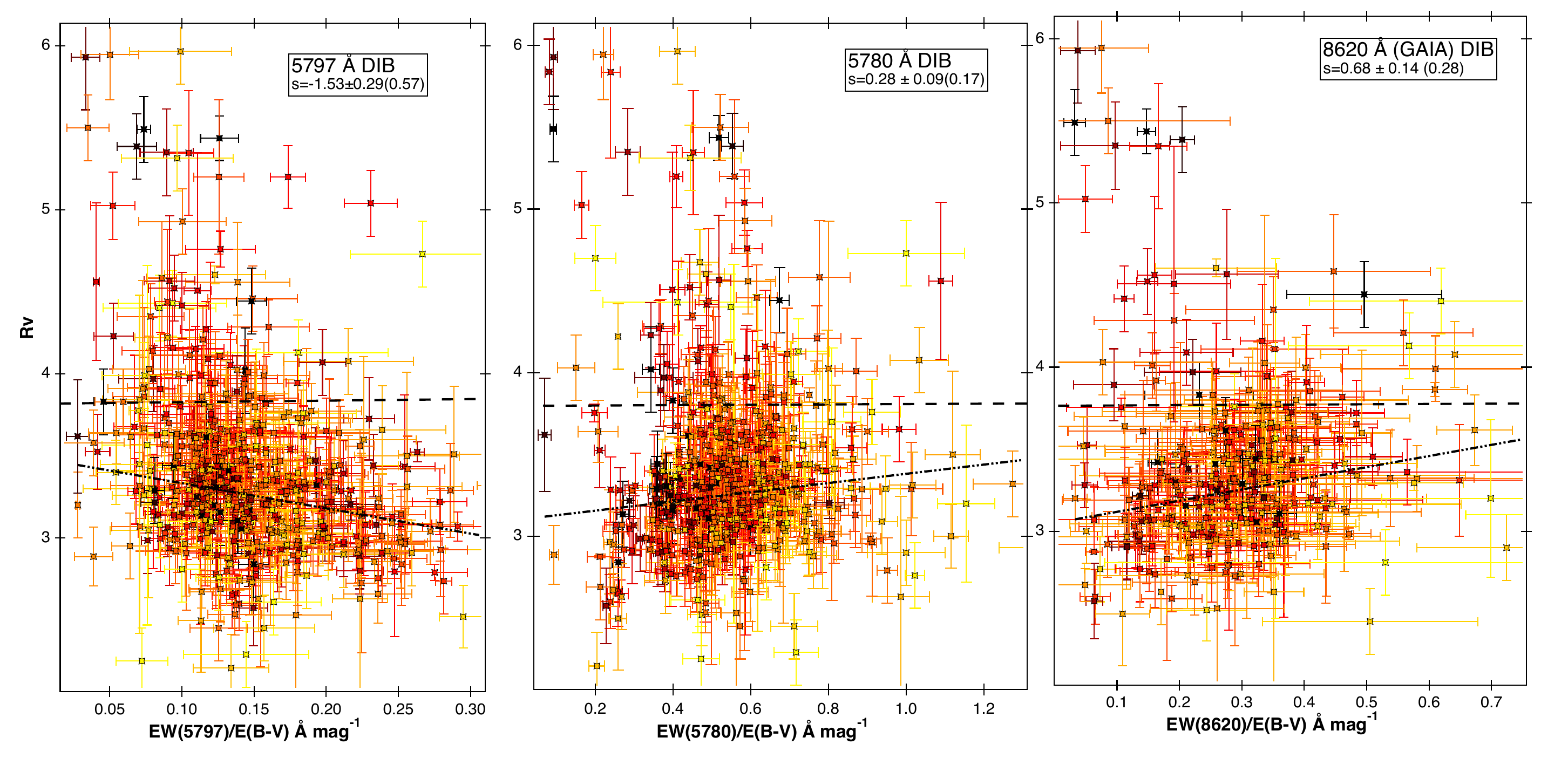}
\caption{Same as Fig. \ref{NORMAV_3DIBS_ALL} for DIB EW's normalized to the reddening E(B-V).} 
\label{NORMEBV_3DIBS_ALL}  
\end{figure*}

Figure \ref{NORMEBV_3DIBS_ALL} is identical to Figure \ref{NORMAV_3DIBS_ALL} except that it uses EW's normalized to the reddening E(B-V) instead of $A_V$. It also shows an evolution among the three DIBs from negative to positive links. A very weak anti-correlation persists for the 5797 \AA~ DIB, while for the two others a positive correlation is definitely visible. We are aware of the common factor 1/E(B-V) in the compared quantities, and the potential positive correlation it may introduce. However, as we could check by comparing all sources, the reddening is quite accurately measured in the case of those distant bright O, B stars, and we do not believe this is the case. Note that in the case of the use of DIB EW ratios discussed below, extinction or reddening values do not play any role. The observed difference among the DIBs confirms the evolution of links between DIBs and grain distributions. As a consequence of all relationships, it appears that the $\zeta$-type 5797 \AA~ DIB is the most appropriate in terms of diagnostic of very high $R_\mathrm{V}$ values. When its strength EW/$A_V$ (resp. EW/ E(B-V)) is below $\simeq$ 0.03  \AA~ mag$^{-1}$ (resp. $\simeq$ 0.13  \AA~ mag$^{-1}$), $R_\mathrm{V}$ has a high probability to be in the very high regime. In the case of the two other DIBs, a net threshold for high $R_\mathrm{V}$ based on the DIB strength does exist only for EW/$A_V$. 

\begin{figure*}
\includegraphics[width=0.98\textwidth,height=8 cm]{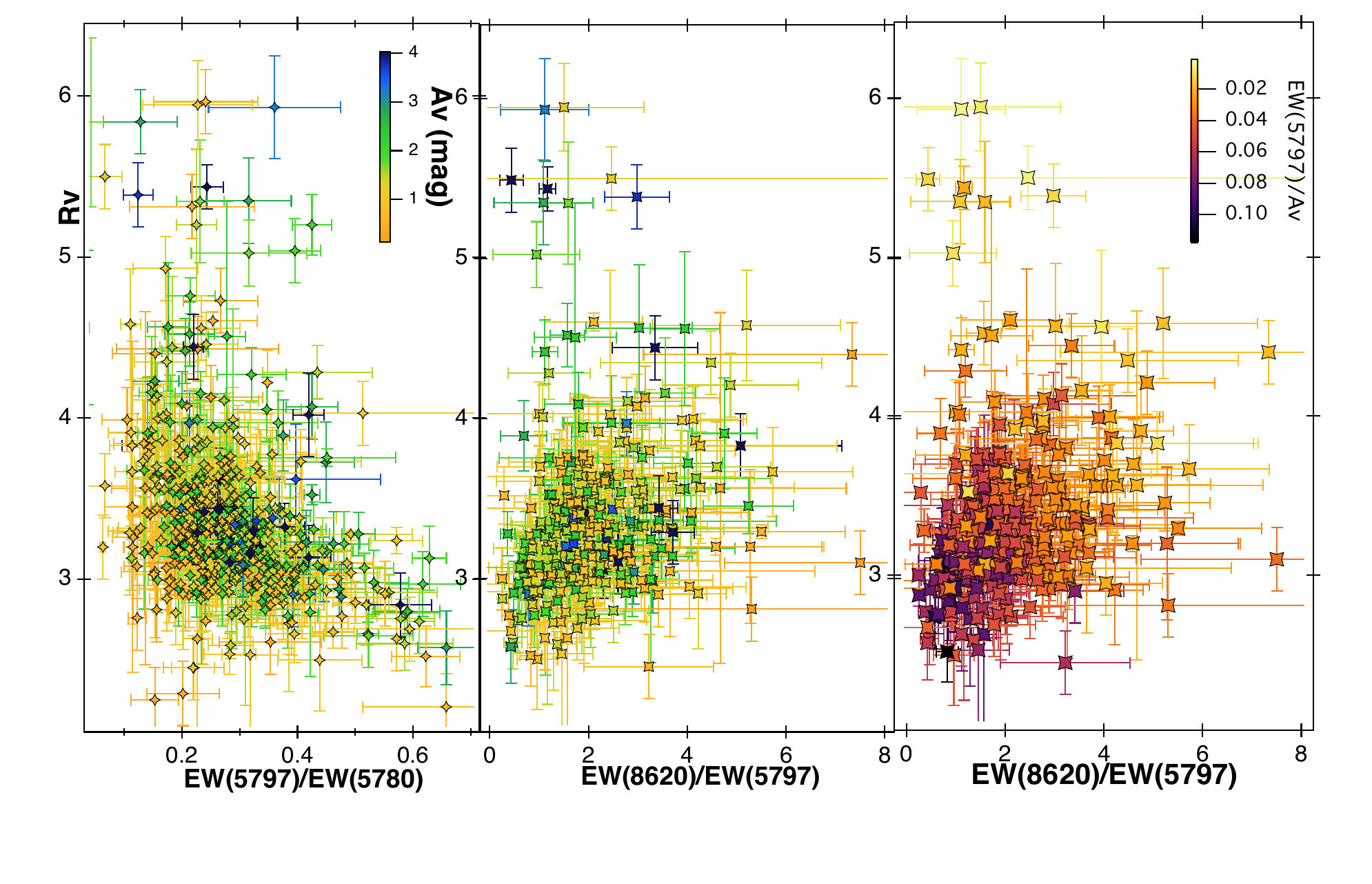}
\caption{Left: Average $R_\mathrm{V}$ from individual LOS vs the 5797 to 5780 DIB strength ratio. The color code indicates the LOS extinction. Middle: same as left for the 8620 to 5797 ratio. Right: same as middle except for the color coding, here showing the extinction-normalized 5797 \AA~ EW. The  combination of the position in the graph and this quantity allows to constrain the $R_\mathrm{V}$  domain, especially high $R_\mathrm{V}$ 's.}
\label{RSUM_RATIOS}
\end{figure*}

\begin{figure}
\includegraphics[width=0.4\textwidth, height=4.5cm]{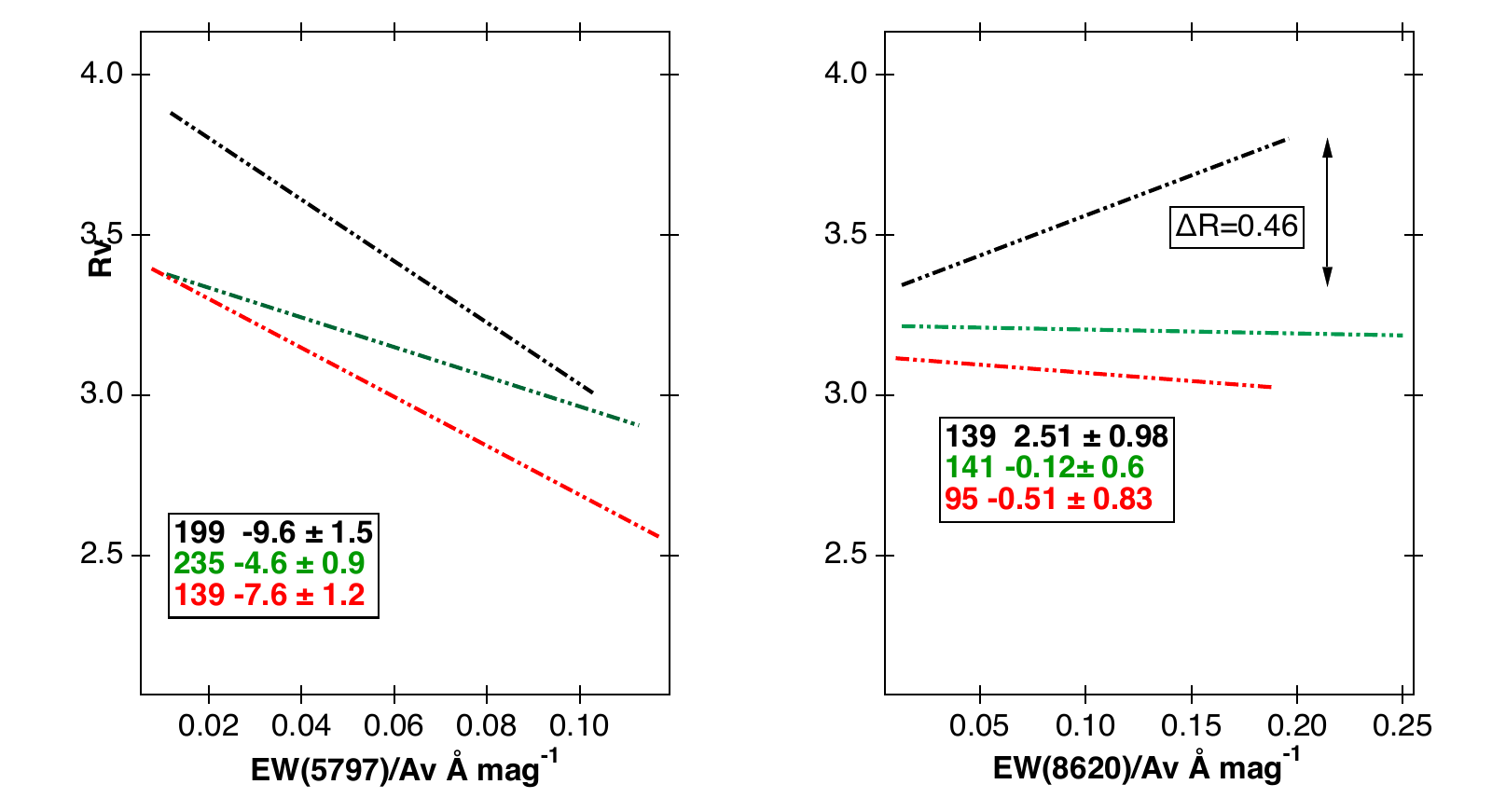}
\caption{Linear fits to $R_\mathrm{V}$ as a function of the 5797 \AA~ DIB (left) and the Gaia DIB (right) normalized to the extinction. $R_\mathrm{V}$ data from \cite{Maiz14} (black), \cite{Wegner03} (green), \cite{Fitzpatrick2007} (red). In the case of the Gaia DIB, there is a change of sign of the relationship. The full $R_\mathrm{V}$ increase in the case of the Gaia DIB and \cite{Maiz14} determinations is indicated, to compare with Fig. \ref{R55GAIAMAPS} (see text).} 
\label{EXPLAIN_DIFF_MAP_LOS_GAIA}  
\end{figure}

\begin{figure}
\includegraphics[width=0.48\textwidth,height=5cm]{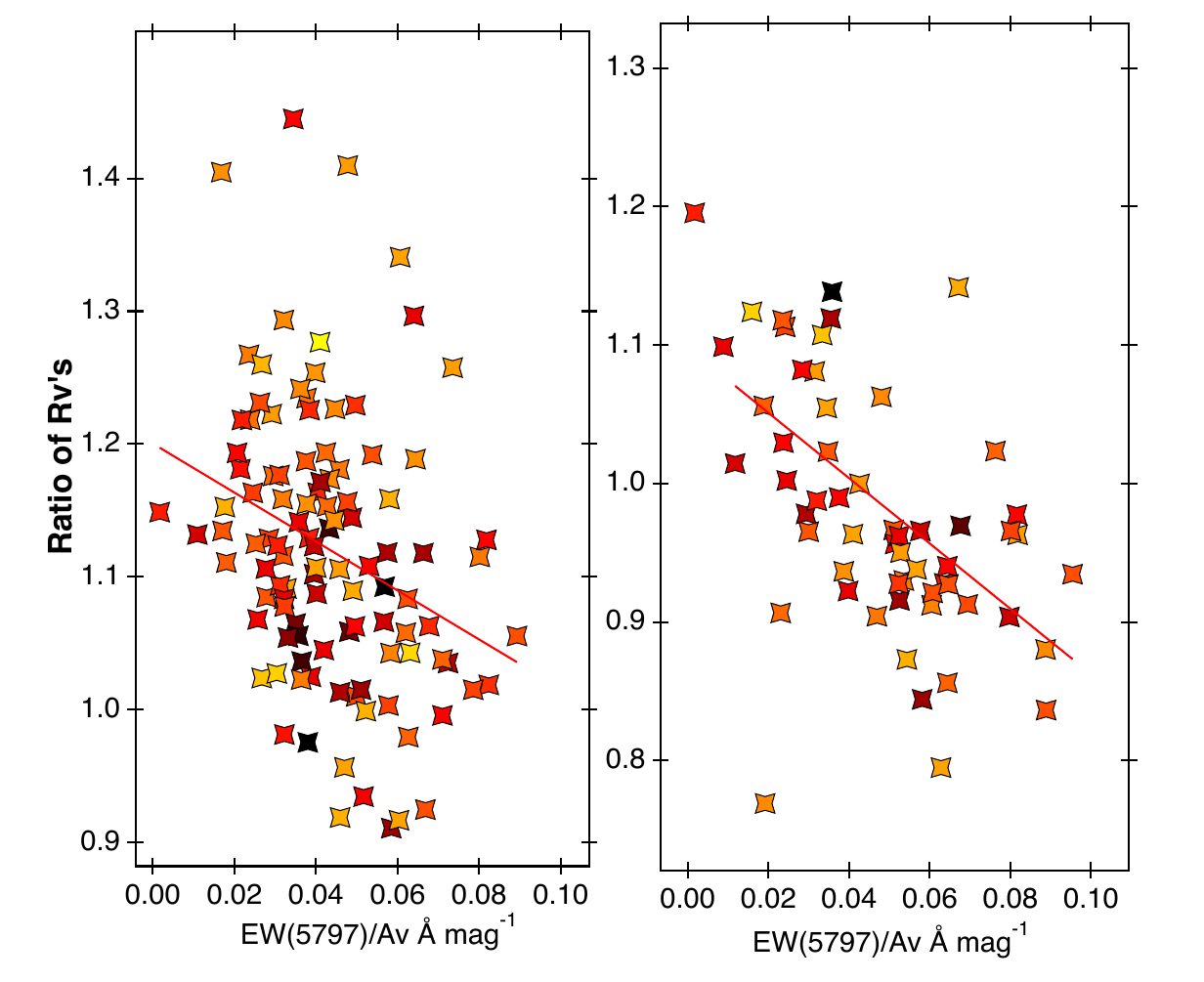}
\caption{DIBs and influence of $R_\mathrm{V}$ determination method. Left: Ratio between the $R_\mathrm{V}$ value from \cite{Maiz14} and the one of \cite{Wegner03} for stars in common, as a function of the 5797 \AA~ DIB normalized to the extinction. Right: Same as left for the ratio between the $R_\mathrm{V}$ value from \cite{FitzpatrickMassa07} and the one of \cite{Wegner03}. The two very low values marked in Fig. \ref{graph_allrvs} have been discarded. Linear fits are displayed. Using the dispersion of the data points around the linear relationships as the average error on $R_\mathrm{V}$, the slopes and 1$\sigma$ errors on slopes are -1.84 $\pm$ 0.57 and -2.13 $\pm$ 0.48 respectively.}
\label{RSUM_EWMH2_NORMAV_GOSS_FM07_WE03}
\end{figure}

Because of the differences observed among DIBs regarding the relationship between $R_\mathrm{V}$ and the DIB strength, the ratios of the EWs of two different DIBs can provide useful information about $R_\mathrm{V}$. Figure \ref{RSUM_RATIOS} displays $R_\mathrm{V}$ as a function of the ratios between the EW's of the Gaia DIB and the 5797 \AA~ DIB, and between those of the 5797 \AA~ and 5780 \AA~ DIBs, revealing a positive and then a negative correlation. There are very strong dispersions of the data points, as discussed above, and better measurements and cloud disentangling are needed, however, such ratios have the advantage of being made from direct measurements, free of calibration, and of being fully independent of extinction or reddening measurements. Still, none of these ratios alone allows to predict very high $R_\mathrm{V}$ values. To do so, however, one may use a combination of DIB ratio and DIB strength. The right panel of Fig. \ref{RSUM_RATIOS} is identical to the middle panel except for the color coding now showing the EW of the 5797 \AA~ DIB normalized to the extinction. Very high $R_\mathrm{V}$'s require simultaneously a low Gaia to 5797 \AA~ DIB ratio and a low 5797 \AA~  DIB strength, which removes the ambiguity. 

 Instead of using a unique value of $R_\mathrm{V}$ averaged over all datasets, we distinguished individual sources. Figure \ref{EXPLAIN_DIFF_MAP_LOS_GAIA} displays the fitted linear relationships between $R_\mathrm{V}$  and the 5797 (resp. 8620) \AA~ DIB strengths, again limited to data points with $R_\mathrm{V}\leq3.8$,  for the three catalogs of \cite{Wegner03}, \cite{Fitzpatrick2007}, \cite{Maiz14}. In the case of the monochromatic R(4595) of \cite{Maiz14}, there is a positive correlation with the Gaia DIB strength and the slope is similar to the one found using the ZG25 determination. On the other hand, the correlation with the Gaia DIB is null or negative in the case of the measurements of \cite{Wegner03} and \cite{Fitzpatrick2007}. This explains why the average value of $R_\mathrm{V}$ shown in Fig. \ref{NORMAV_3DIBS_ALL}, a combination of all measurements, appears independent of the Gaia DIB. Figure \ref{EXPLAIN_DIFF_MAP_LOS_GAIA} also displays the variation of the $R_\mathrm{V}$ value for the same three $R_\mathrm{V}$ datasets as a function of the 5797 \AA~ DIB strength. It can be seen that all correlations are negative, which explains the negative trend found for the average value. 
 
 In our previous comparison shown in Figure \ref{EXPLAIN_DIFF_MAP_LOS_GAIA}, we have considered separately the datasets associated with the different methods of $R_\mathrm{V}$ measurement,  i.e. datasets made of different target lists, with potential biases linked to the target selection.  In order to check unambiguously the influence of the method of $R_\mathrm{V}$ measurement on the link with the DIBs and avoid any bias, we selected the group of stars for which the two $R_\mathrm{V}$ determinations do exist.  
Figure \ref{RSUM_EWMH2_NORMAV_GOSS_FM07_WE03} displays the ratio between $R_\mathrm{V}$s obtained from two different methods as a function of the 5797 \AA~ DIB strength.  The figure shows that there is a small and marginally detectable decrease of the ratio between the $R_\mathrm{V}$ value based on the \cite{Maiz14} method and $R_\mathrm{V}$ derived from the IR extrapolation of the EC used by \cite{Wegner03}. On the other hand, a clearly defined trend is found in the case of the ratio of $R_\mathrm{V}$ based on the \cite{Fitzpatrick2007} method and  $R_\mathrm{V}$ from \cite{Wegner03}. For both types of ratio, we show in the figure  the numerical values of the slope of a linear fit to the data and its estimated errors. We cautiously interpret such trends as due to a different link between the DIB carrier abundance and each of the three wavelength ranges of the EC influencing at most the $R_\mathrm{V}$ value for each technique. This is in agreement with a dependence of the abundance of a given DIB to a specific wavelength range in the EC. 

\section{Summary and discussion}

We have built and/or assembled datasets allowing comparisons between the $R_\mathrm{V}$ parameter of the extinction curve and the strength of three different DIBs. We compiled published $R_\mathrm{V}$ determinations for 505 O-, B-type target stars into a single catalog, including multiple estimates for the same target and the corresponding sources, and an average of the various estimates. We searched for high-resolution spectra of those stars in public archives, fitted the continua around the 5780\AA~, 5797\AA~ and 8620\AA~ (Gaia) DIBs and extracted the EWs in a classical way. For 91 stars we used the  5780\AA~ and 5797\AA~ compiled values of \cite{Xiang17}. $R_\mathrm{V}$s and DIB data were compared.  In the case of the 8620 \AA~ (Gaia) DIB, we also used previously published 3D maps of the DIB, and made comparisons with 3D maps of both local and LOS-integrated $R_\mathrm{V}$ from \cite{ZhangGreen25} (ZG25).  Additionally, for the purpose of the study, a new 3D extinction map has been produced, by inversion of the extensive catalog of extinction data for individual stars from \cite{Zhang23}, and is made available.

\begin{itemize}
    \item 
    We find new evidence for large spatial scale variations of $R_\mathrm{V}$, based on the compilation of $R_\mathrm{V}$s from hot stars.  To do so, we inverted the \cite{Zhang23} catalog of extinction data based on Gaia XP spectra, then used the resulting 3D extinction map to identify, along each LOS to a hot star, those segments of high extinction density responsible for the LOS-integrated $R_\mathrm{V}$ value.  The combination of those segments and their $R_\mathrm{V}$ values provides a partial mapping of $R_\mathrm{V}$, which shares common large-scale features with the ZG25 map, despite fully independent sources.   E.g., the Sagittarius-Carina (Sag-Car) clouds have a higher $R_\mathrm{V}$ than those that make up the Split cloud group. ZG25 attribute this effect to the large number of young stars present in Sag-Car and their destructive effects on small grains. An examination of the distribution of massive O- and B-type stars, e.g. based on the \cite{Pantaleoni21} catalog \citep[see, e.g. their distribution superimposed on an extinction map in][]{Lallement22}, and the comparison with the distribution of $R_\mathrm{V}$, further supports  this hypothesis. 

\item We show that there is a striking similarity between the distribution of local $R_\mathrm{V}$ shown by ZG25 and the distribution of the Gaia DIB abundance relative to the extinction density (Fig. \ref {Large_scale_rv_dibtoext}), with high $R_\mathrm{V}$'s corresponding with high DIB abundance. 


\item We also find a global positive correlation between LOS-integrated R55 values from the catalog of ZG25 and Gaia DIB strengths,  in agreement with the spatial similarity. The Gaia DIB strengths in this case were obtained by integrating within the \cite{Cox24}  3D DIB and extinction maps along all LOS from the catalog. The positive correlation ends for high DIB abundances, i.e. for LOS crossing only very diffuse ISM regions, and there is a flattening at R55 $\simeq$3.0.  In the domain of positive correlation, R55 is lower where the LOS extinction $A_\mathrm{V}$ is higher, a pattern found by ZG25 for the nearby clouds. A more detailed analysis shows that along with the global positive correlation there are complex effects associated with the LOS that form the wide bump responsible for the observed curvature and the very high R55 points on top of it. We interpret this bump as the consequence of the presence of dense clouds with a high local $R_\mathrm{V}$ along a fraction of those LOS, and the resulting increase of the LOS-averaged $R_\mathrm{V}$. The observed skewed distribution of $R_\mathrm{V}$'s favors our interpretation. The lack of link with the DIB for this population is due to the absence of the Gaia DIB carrier in the localized dense parts. The same is true for the small conspicuous group of very high $R_\mathrm{V}$'s above the bump, corresponding this time to short sight lines (less than several hundred parsecs) easily identified as those crossing nearby dense clouds. This shows that, despite a global positive correlation, the Gaia DIB strength alone can not be a tracer of the presence of opaque clouds with high $R_V$ along distant lines of sight and may only be a proxy for $R_\mathrm{V}$ for the diffuse ISM. 
\item 
The study  of the DIB-$R_\mathrm{V}$ relationships for the individual LOS to hot stars reveals a clear differentiation among the three DIBs, with, especially,  a marked difference between the $\zeta$-type 5797 \AA~ DIB and the two other $\sigma$-type DIBs. $R_\mathrm{V}$ is monotonically decreasing with the 5797 \AA~ DIB strength, making this DIB a potential diagnostic of high and low $R_\mathrm{V}$ LOS. This anti-correlation weakens strongly for the 5780 \AA~ DIB then turns to a barely visible positive correlation for the Gaia DIB (affected by the large error bars  on the Gaia DIB EWs). The sign reversal from the 5797 \AA~ DIB to the 8620 \AA~ DIB is more clearly visible if one uses the DIB EW normalized to the reddening (Fig. \ref{NORMEBV_3DIBS_ALL}). As a consequence of these opposite relationships, a potential diagnostic of $R_\mathrm{V}$ is the ratio between 5797 \AA~ and 5780 \AA~ EWs, or between 5797 \AA~ and 8620 \AA~ EWs, which have both the advantage of being direct calibration-free measurements and do not require additional reddening or extinction estimates. On the other hand, predictions of very high $R_\mathrm{V}$'s require a simultaneously use of the Gaia to 5797 \AA DIB ratio and the 5797 \AA~  DIB strength, and low values of both quantities.

\item 
An unexpected result, calling for further study,  is the way the $R_\mathrm{V}$-DIB relationship depends on the method used for the determination of $R_\mathrm{V}$. In the case of the Gaia DIB, different methods may result in opposite signs of the DIB strength-$R_\mathrm{V}$ relationship, which explains why an average over all $R_\mathrm{V}$ measurements does not reveal the positive correlation found based on 3D maps and the ZG25 determination, while an agreement is found when using the spectroscopic R(5495)= A(5495)/E(4405-5495)  determination of \cite{Maiz14} alone. In the case of the 5797 \AA~ DIB, its strength is sensitive to the ratio between the $R_\mathrm{V}$ from \cite{Fitzpatrick2007} or R(5495) from \cite{Maiz14} and the \citep{Wegner03} $R_\mathrm{V}$ defined as the intercept of the EC in the FIR. Correlations are found in both cases, with a higher confidence in the case of the former comparison. Our interpretation is that the DIB-grain link is different depending on the wavelength range of the EC influencing the most the $R_\mathrm{V}$ measurement.

\end{itemize}

The various results of this study reveal a large complexity of the links between the properties of the grains shaping the EC and the abundances of DIB carriers. On the other hand, a coherent pattern is also apparent if one considers that each DIB carrier has its own response to the physical properties and environmental conditions of the medium, which are also influencing the grains. In particular, the strongest differences found here between the 5797  \AA~  DIB, known for remaining strong in dense gas, and disappearing only in high density cores, and the two other DIBs, mainly associated with the translucent or diffuse clouds, are very likely related to such characteristics, because each phase corresponds to a different size distribution of grains and a different local EC.  The influence of the $R_\mathrm{V}$ determination technique on the DIB-$R_\mathrm{V}$ link can also be understood if the EC wavelength range with the strongest influence on $R_\mathrm{V}$ also corresponds to a range of grain sizes and the associated conditions, and in turn to a range of abundances of the carriers. A subsequent question is whether or not DIB carriers do participate to the dust chemistry, and how. In this respect, the opposite signs of the DIB-$R_\mathrm{V}$ correlations for the 5797  \AA~  and Gaia DIBs enter in resonance with the recent findings of ZG25, namely the existence of a bi-modal dependence of $R_\mathrm{V}$ on the extinction. According to the interpretation of the authors, from low to moderate extinctions, there is a first phase of accretion onto ultra-small grains and a subsequent increased fraction of the absorption by small grains, resulting on a decrease of $R_\mathrm{V}$. At higher opacity starts a second phase of coagulation, forming large grains and responsible for $R_\mathrm{V}$ increase to very high values. The positive (resp. negative) correlation between $R_\mathrm{V}$ and the Gaia (resp. 5797  \AA~ ) DIB may be a sign of disappearance of the DIB carriers associated with the first accretion phase (resp. second coagulation phase) conditions. In other words, $\sigma$-type DIBs such like the Gaia DIB start disappearing in regions of low-moderate extinction when $R_\mathrm{V}$ is decreasing, while $\zeta$-type DIBs as the 5797  \AA~ DIB start decreasing in dense molecular regions, where $R_\mathrm{V}$ is increasing. This suggests that carriers of DIB are at least a fraction of the carbonaceous  matter coating the grains \citep[see, e.g., models by][]{Jones2014}.



In future, more precise and detailed relationships between DIBs and the EC are expected in several ways. High resolution stellar surveys with existing or future multi-object spectrographs are expected to provide massive measurements of DIB EW's, not only for hot stars but for stars of all types in case of the use of stellar models \citep[e.g.][]{Kos13, Puspitarini15, Chen13coolstars}.  DIBs are weak features but present the strong advantage of being direct measurements totally free of any calibration. Moreover, being cumulative quantities along lines of sight, it is relatively easy to invert their EW's to build detailed 3D maps. 3D maps of DIBs other than the Gaia DIB would allow to extend correlations with the EC and refine the diagnostics. The normalization to the extinction will also be facilitated by the future 3D extinction maps expected from the next Gaia data releases. Doppler shifts measurements of DIBs may also help redistributing the carriers in the different clouds located along the paths to target stars \citep[see, e.g. the kinematical studies based on DIBs by][]{Zasowski15, Puspitarini15}. Finally, of particular interest would be the use of DIBs associated to dense phase (e.g. the 5850 \AA~  DIB, a $\zeta$ prototype), or one of the intermediate DIBs \citep{Omont24}. According to the results for the 5797  \AA~ DIB,  their relationship with $R_\mathrm{V}$ may be even tighter. In parallel, identifying DIB carriers through comparison with laboratory studies will, we hope, lead to concrete progress in this field and shed light on the chemical processes involving these carriers in the various phases of the interstellar medium (ISM), as well as on their relationship with dust grains.

\begin{acknowledgements}

We thank our referee for constructive comments, which helped us clarify our presentation. Based on data obtained from the ESO Science Archive Facility, including observations made with ESO Telescopes at the La Silla and Paranal Observatories. Based on data obtained from the POLARBASE Science Archive Facility, including observations obtained at the Canada-France-Hawaii Telescope (CFHT) which is operated by the National Research Council of Canada, the Institut National des Sciences de l'Univers of the Centre National de la Recherche Scientifique of France, and the University of Hawaii, and observations obtained at the Telescope Bernard Lyot (USR5026) operated by the Observatoire Midi-Pyrénées, Université de Toulouse (Paul Sabatier), Centre National de la Recherche Scientifique of France. 
This work has made use of data from the European Space Agency (ESA) mission Gaia (https://www.cosmos.esa.int/gaia), processed by the Gaia Data Processing and Analysis Consortium (DPAC, https://www.cosmos.esa.int/web/gaia/dpac/consortium). Funding for the DPAC has been provided by national institutions, in particular the institutions participating in the Gaia Multilateral Agreement. 
This research has made use of the VizieR catalog access tool, CDS, Strasbourg, France (DOI: 10.26093/cds/vizier). The original description of the VizieR service is published in 2000, A\&AS 143, 23.

\end{acknowledgements}

\bibliographystyle{aa}
\bibliography{dibgaiaextlaw.bib}

\begin{appendix}

\section{The new 3D extinction map}
\begin{figure}
\includegraphics[width=0.5\textwidth]{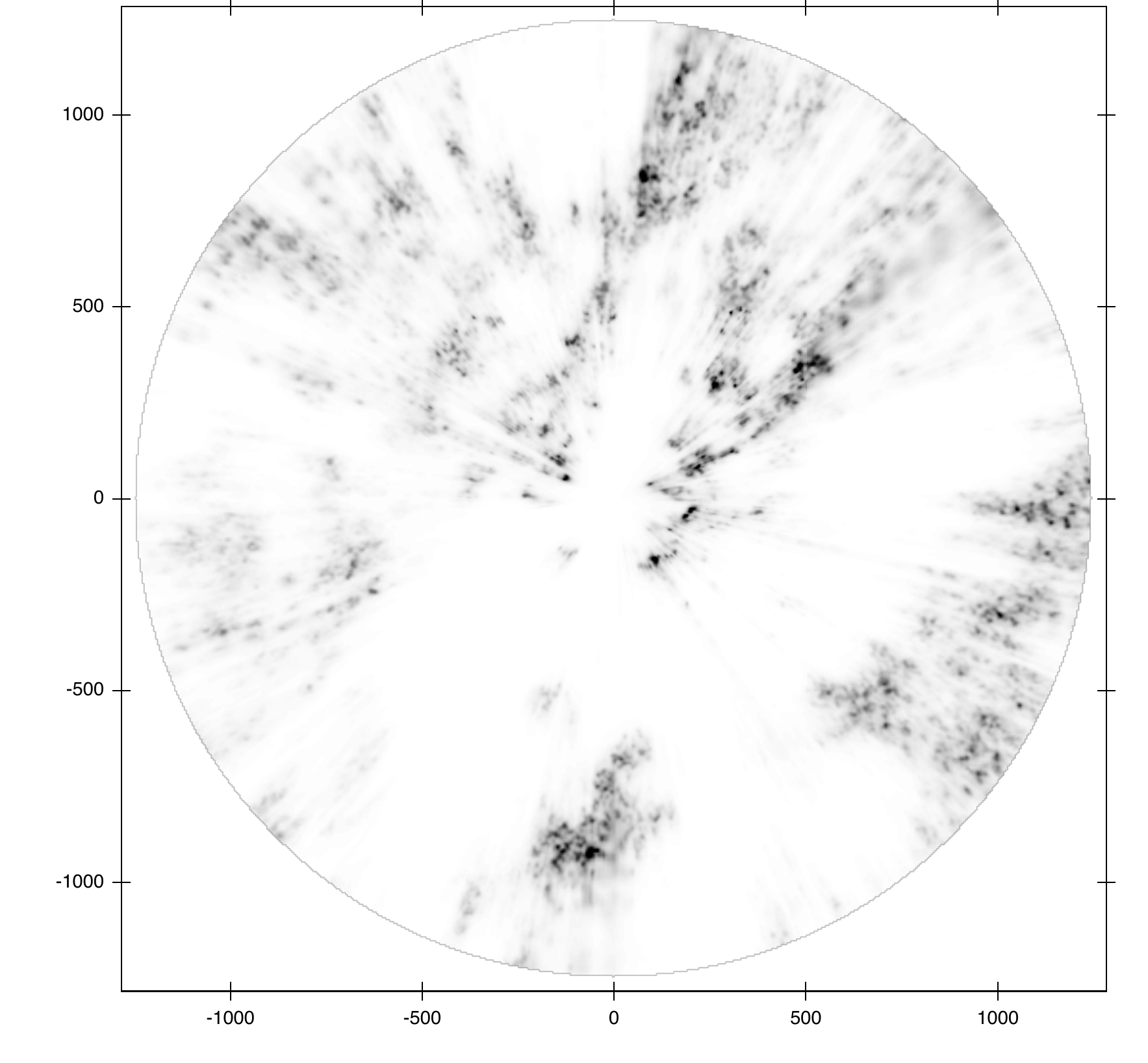}
\includegraphics[width=0.5\textwidth]{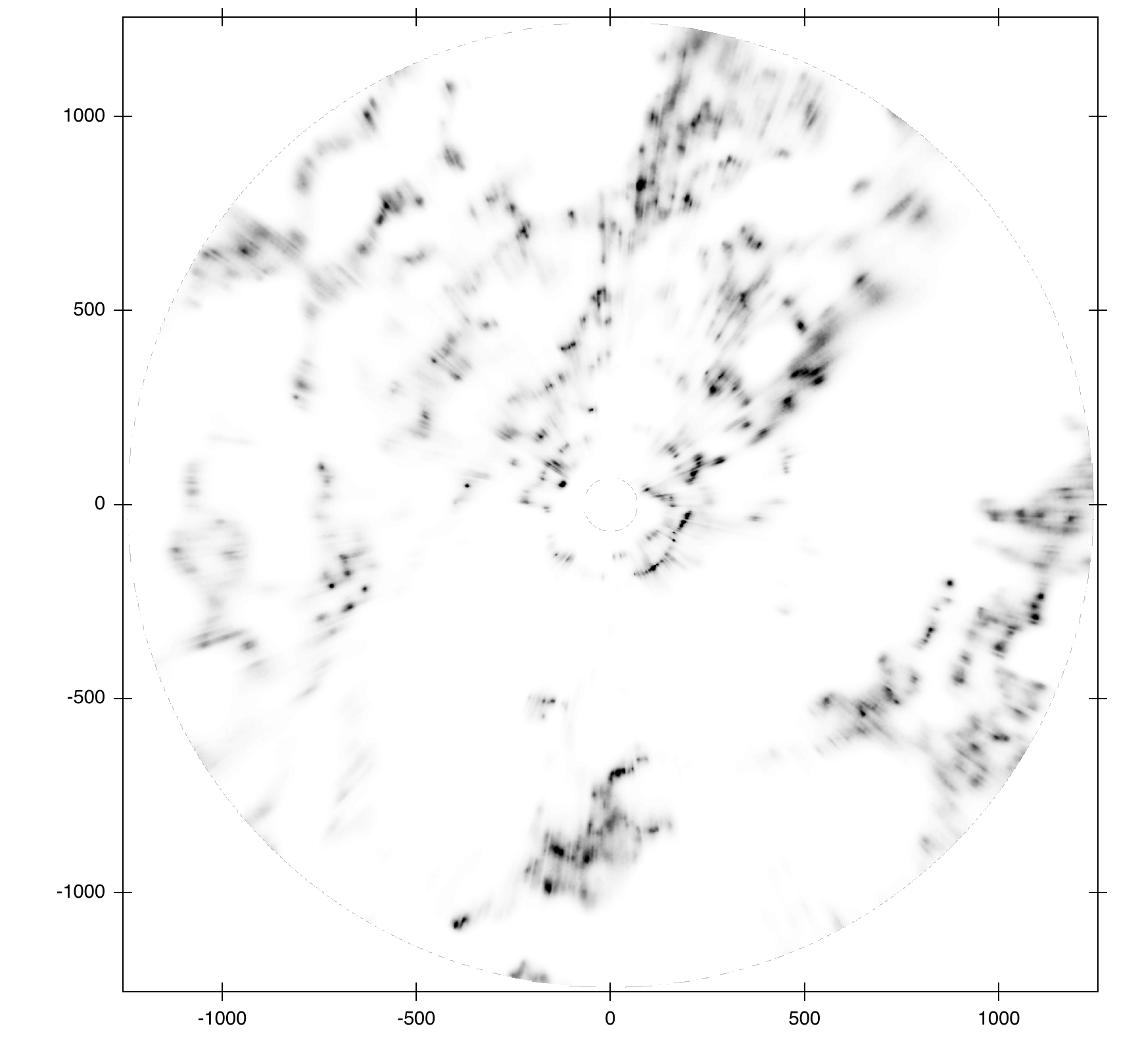}
\caption{Illustration of the properties of our new 3D extinction map based on the hierarchical inversion of the \cite{Zhang23} extinction catalog. Top: the computed extinction density in the z=0 plane, here limited to distances shorter than 1.2 kpc. Bottom: extinction density from the \cite{Edenhofer24} map, also in the z=0 plane. The resolution of our new inversion is between 5 and 10 pc depending on the location, poorer than the high, 1 pc resolution achieved by \cite{Edenhofer24}. This explains why clouds appear wider and more diffuse in the top graph. Still, all structures are very similar. The full inversion (not shown here) has produced new maps extending up to 10 kpc from the Sun and is available with tools on the Datalabs platform. The resolution is adapted to the density of target stars, i.e. the resolution gets poorer at large distance due to the lack of target with accurate parallaxes and photometric data. The map has been used to find the high extinction parts along the path to the 505 target stars of sections 3 to 5. }
\label{comp3dmaps_10_and_5}
\end{figure}

Figure \ref{comp3dmaps_10_and_5} shows a planar section along the Galactic plane in the 3D map of extinction density briefly described  in section \ref{3D}. The map is based on the catalog of \textbf{LOS} extinction estimates of \cite{Zhang23}. The \textbf{inversion} technique is identical to the one used for the 3D map of \cite{Vergely22}. Details can be found in \cite{Lallement22} and \cite{Vergely22}. The main characteristic of the technique is the limitation of the map resolution to the density of available target stars everywhere in space, by means of an iterative, hierarchical approach. There is also an imposed minimum size of the structures. 4 different maps have been computed, corresponding to different volumes and minimum sizes. The smallest map is limited to the first 500 pc. Its maximum resolution is  5pc and is reached in most regions. The second map is reaching 1500 pc and has a maximum resolution of 10 pc. The third map is reaching 4kpc and the maximum resolution is 25pc. The widest map reaches 10 kpc and has a maximum resolution of 50pc.  The full map extends to 10 kpc and is available online with appropriate tools. In figure \ref{comp3dmaps_10_and_5}, the 5 pc and 10 pc resolution maps have been used together, and the wider 10 pc resolution map has been limited to a radius of 1.2 kpc for comparison  with the very high resolution map of \cite{Edenhofer24}, based on the same input data. 
The comparison shows that all dust concentrations present in the high resolution map of \cite{Edenhofer24} are also retrieved in our inversion, although broadened due to our larger spatial resolution (here 5 pc or 10 pc depending on the locations). The four maps and associated tools will be available on the Datalabs platform \footnote{\url{https://datalabs.esa.int/}} or at the ACRI-ST address \footnote{\url{ https://gtomo.dev.acrist-services.com/}} . The four maps will be merged into a unique map taking advantage of the best resolution everywhere, using a variable distance step.

\newpage

\section{DIB equivalent width measurements}

\begin{figure}
\includegraphics[width=0.45\textwidth,height=4.5cm]{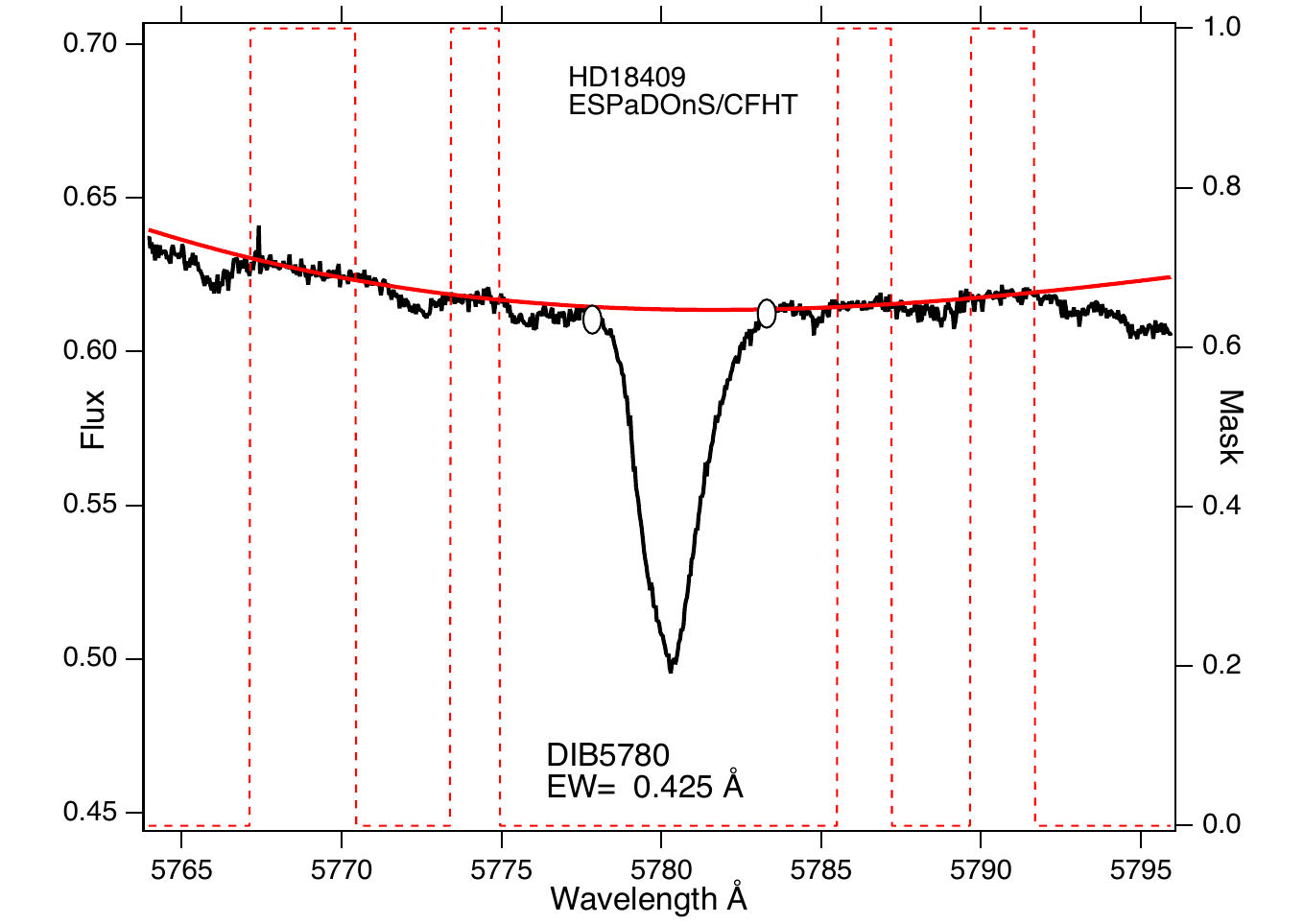}
\includegraphics[width=0.45\textwidth,height=4.5cm]{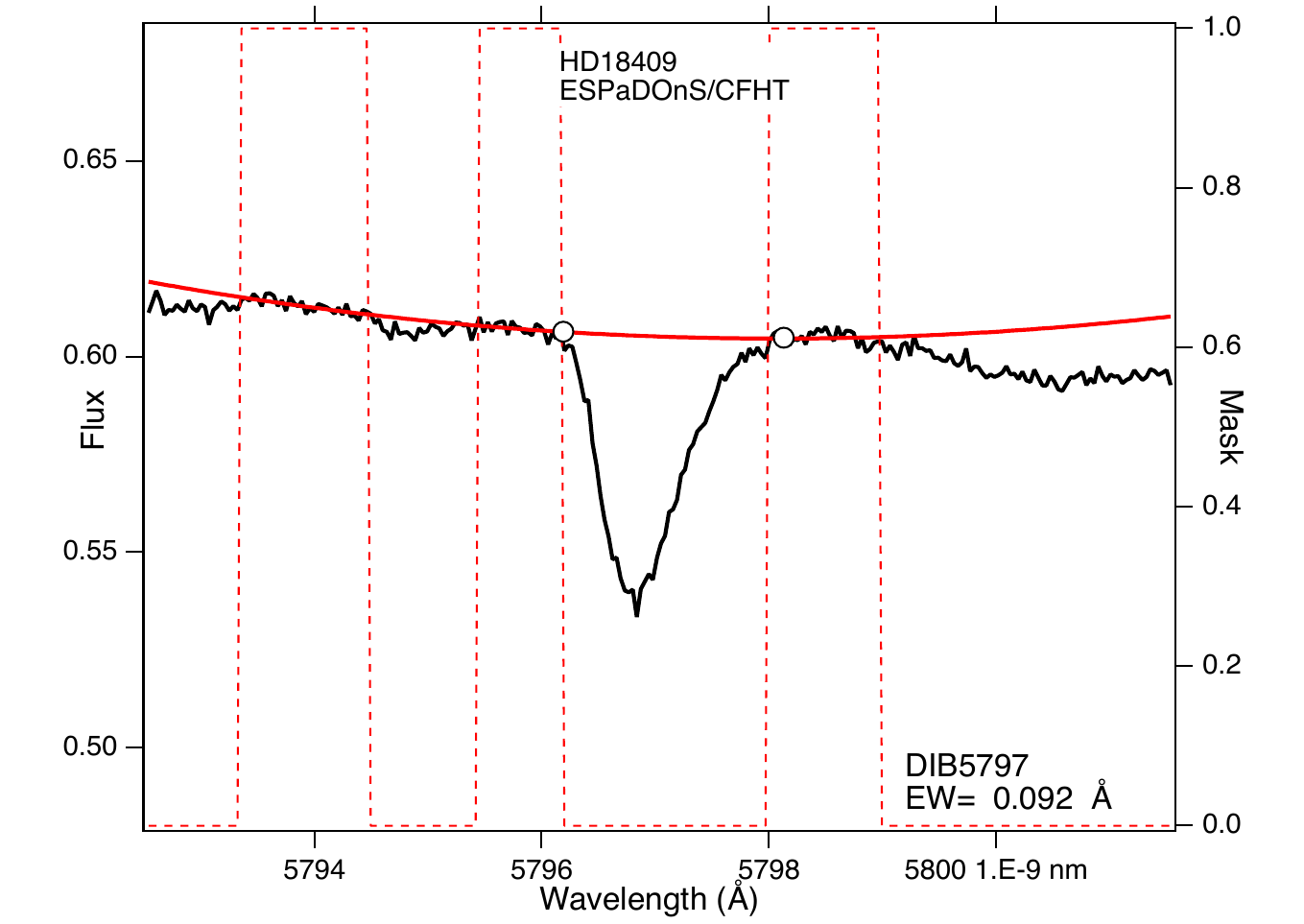}
\includegraphics[width=0.45\textwidth,height=4.5cm]{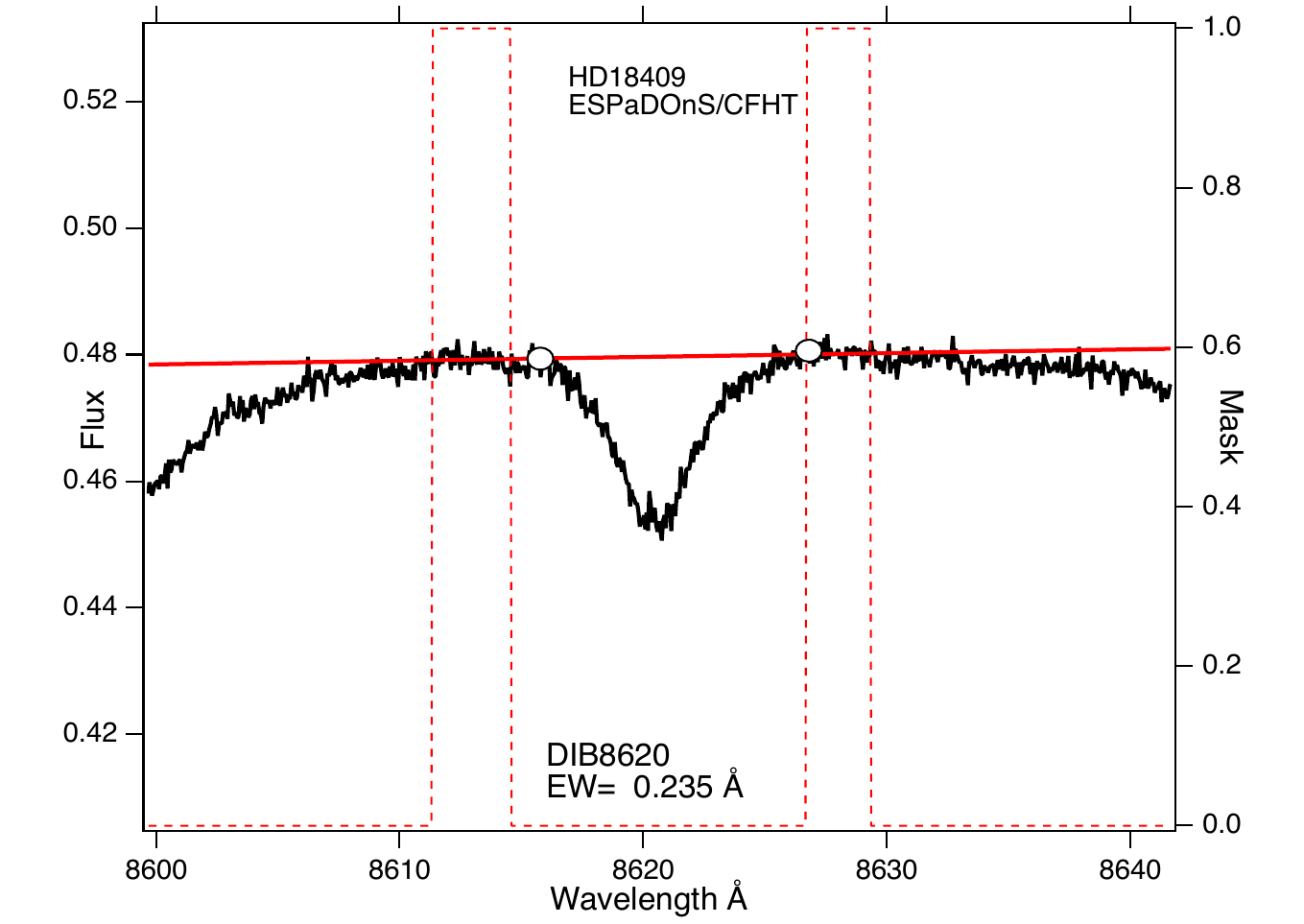}
\caption{Examples of DIB EW measurements, here for the $A_V$=2.2 mag  O star HD18409 from the GOSSS survey of \cite{Maiz18}. Data (black) from the ESPAdOns spectrograph are fitted to the product of a polynomial (solid red) in the unmasked wavelength intervals (dashed red line). After division by this fitted continuum, the DIB EW is computed between the two marked cursors (empty circles).}
\label{dib-fits}
\end{figure}

Figure \ref{dib-fits} displays examples of EW measurements for the three different DIBs considered in this work. The EW's are obtained in the classical way described in section \ref{dib_extraction}, along with the specific characteristics of the measurements with the different instruments.

\end{appendix}
\end{document}